\documentclass[aps,prb,twocolumn,superscriptaddress,showpacs]{revtex4-1}

\usepackage[dvips]{graphicx}
\usepackage{color}
\usepackage{latexsym}
\usepackage{amsmath}
\usepackage{amsthm}
\usepackage{amsfonts}
\usepackage{amssymb}
\usepackage{mathptmx}
\usepackage{subfigure}
\usepackage{cancel}

\providecommand*{\unit}[1]{\,\ifmmode \mathrm{\,#1}\else\textup{#1}\fi}

\begin{document}

\title{Effect of Photon-Assisted Andreev Reflection in the Accuracy of a SINIS Turnstile}

\author{A. Di Marco}
\thanks{Current address: Department of Nanotechnology and Nanoscience (MC2), Chalmers University of Technology, SE-41298 G\"{o}teborg, Sweden}
\affiliation{LPMMC-CNRS, Universit{\'e} Grenoble-Alpes, 25 Avenue des Martyrs B.P. 166, 38042 Grenoble Cedex, France}

\author{V. F. Maisi}
\thanks{Current address: Solid State Physics Laboratory, ETH Zurich, 8093 Zurich, Switzerland}
\affiliation{Centre for Metrology and Accreditation (MIKES), P.O. Box 9, 02151 Espoo, Finland} 
\affiliation{Low Temperature Laboratory, Aalto University, P.O. Box 13500, FI-00076 Aalto, Finland}

\author{F.\ W.\ J.\ Hekking}
\affiliation{LPMMC-CNRS, Universit{\'e} Grenoble-Alpes, 25 Avenue des Martyrs B.P. 166, 38042 Grenoble Cedex, France}

\author{J. P. Pekola}
\affiliation{Low Temperature Laboratory, Aalto University, P.O. Box 13500, FI-00076 Aalto, Finland}

\begin{abstract}
We consider a hybrid single-electron transistor (SET) constituted by a gate-controlled normal-metal island (N) connected to two voltage-biased superconducting leads (S) by means of two tunnel junctions (S-I-N-I-S), operated as a turnstile. We show that the exchange of photons between this system and the high-temperature electromagnetic environment where it is embedded enhances Andreev reflection, thereby limiting the single-electron tunneling accuracy.
\end{abstract}


%
%
%
%
%
%
%
%
%
%
%
%
%
%
%
%
%
%

\date{\today}

\maketitle

\section{Introduction}

\begin{figure}[ht!]
\includegraphics[height=2.8cm]{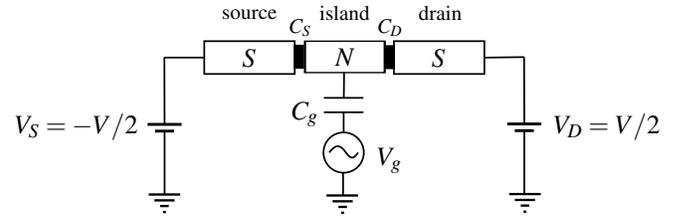}
\caption{Hybrid S-I-N-I-S single-electron transistor (SET). The black parts stand for the insulating barriers of the tunnel junctions.}
\label{fig:circuit}
\end{figure}

The experimental realization of a quantum electric current standard is one of the scientific and technological challenges of the present time. This is a key goal in metrology because it would lead to a modern definition of Ampere as well as to the most accurate comparison of the fundamental constants $R_K=h/e^2$ and $K_J=2e/h$.~\cite{Flowers:2004} Among the devices proposed until now,~\cite{Geerligs:1990,Pothier:1992,Lotkhov:2001,Vartiainen:2007,Blumenthal:2007,Kaestner:2008} the hybrid SINIS single-electron transistor (SET) depicted in Fig.~\ref{fig:circuit} is one of the most interesting candidates.~\cite{Pekola:2008} Such a device is formed of a normal-metal (N) island joined to two superconducting (S) electrodes via two tunnel junctions with capacitances $C_S$ for the source (S) and $C_D$ for the drain (D). The entire structure is biased with a constant voltage $V_D-V_S=V$. The amount of electric charge localized on the island is controlled using a gate potential $V_g(t)$, capacitively coupled to N by means of a gate with capacitance $C_g$. Typically, the charging energy of the island $E_C=e^2/2C_\Sigma$, with $C_\Sigma=C_S+C_D+C_g$,  governs the tunneling processes in the SET, i.e., the system works in the Coulomb blockade regime. Additional energy filtering is provided by the two outer superconductors which protect the device against unwanted tunneling events. In this context, if the single-electron tunneling is the dominant process, a periodic $V_g(t)$ signal with frequency $f$ generates an electric current $I$ through the SET which is equal to $ef$. In other words, the SET is a frequency-to-current converter. However, high-order tunneling events occur in addition to the single-particle ones. They limit the conversion accuracy of this electronic turnstile thereby acting as error sources. The main contribution to the total error is usually provided by elastic and inelastic cotunneling~\cite{AverinNazarov:1990,Averin:1997} as well as Andreev reflection and Cooper-pair cotunneling.~\cite{Averin:2008,Aref:2011} From the theoretical point of view, it has been shown that all these processes can be eliminated efficiently thereby reaching the metrological requirements.~\cite{Averin:2008} Nevertheless, in real experiments the achievement of the accuracy needed for the completion of the so-called quantum metrological triangle remains a difficult task. In particular, a noise-induced residual Andreev tunneling current affects the I-V characteristic of the SET turnstile although the increase of the charging energy $E_C$, with respect to the gap parameter $\Delta$ of the superconductors, leads to a decrease of Andreev reflection probability.~\cite{Aref:2011} Such a two-electron current may be due to the effect of the high-temperature electromagnetic environment the SINIS device is coupled with. The energy provided by such an external thermal bath to the SET via the exchange of photons can promote tunneling of particles through the single junction.~\cite{Pekola:2010,DiMarco,Bubanja:2014} In this paper, we show that, indeed, the environment-assisted Andreev reflection limits the turnstile accuracy, unless it is properly taken care of.

\section{Electronic transport in a SINIS turnstile}

\begin{figure*}[ht!]
\centering  \subfigure[]
{\includegraphics[height=3.9cm]{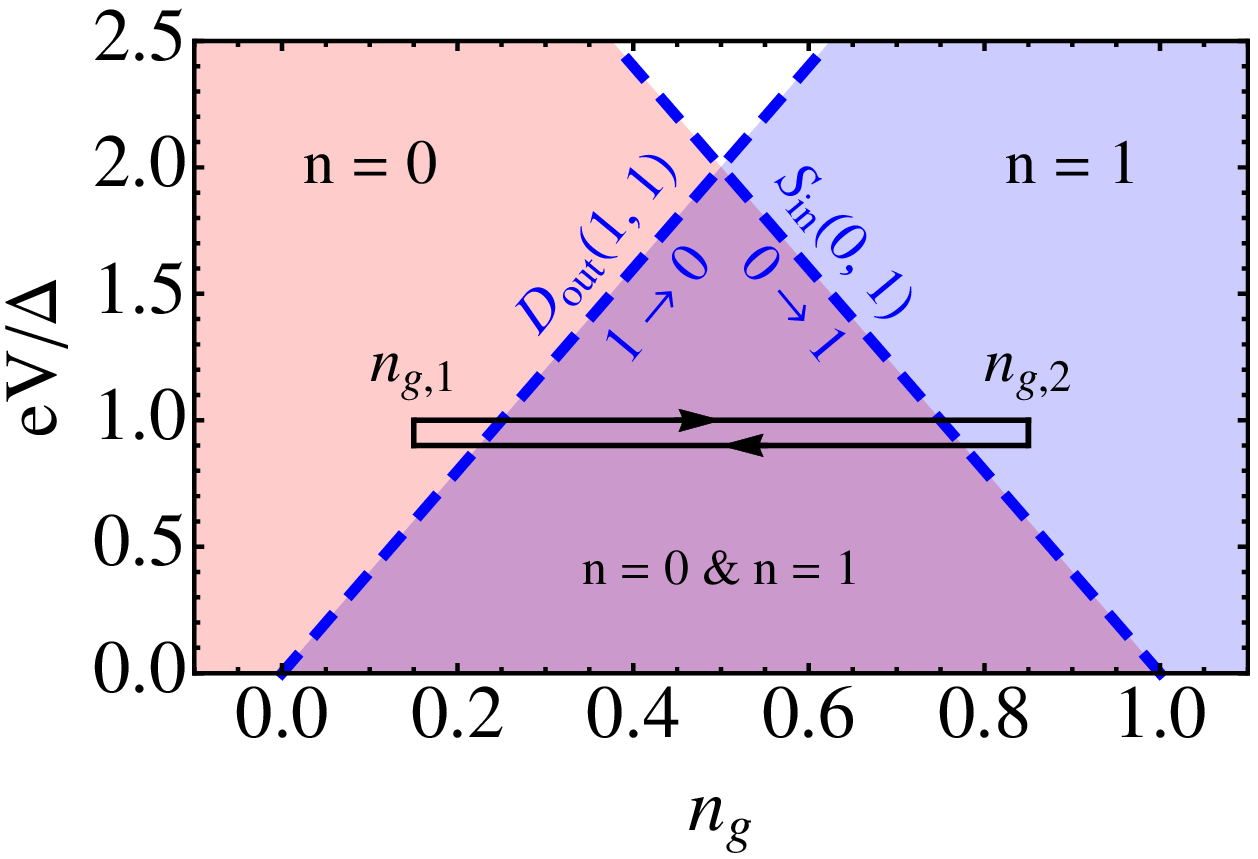}\label{fig:diamonds_single}}
\hspace{3mm} \subfigure[]
{\includegraphics[height=3.9cm]{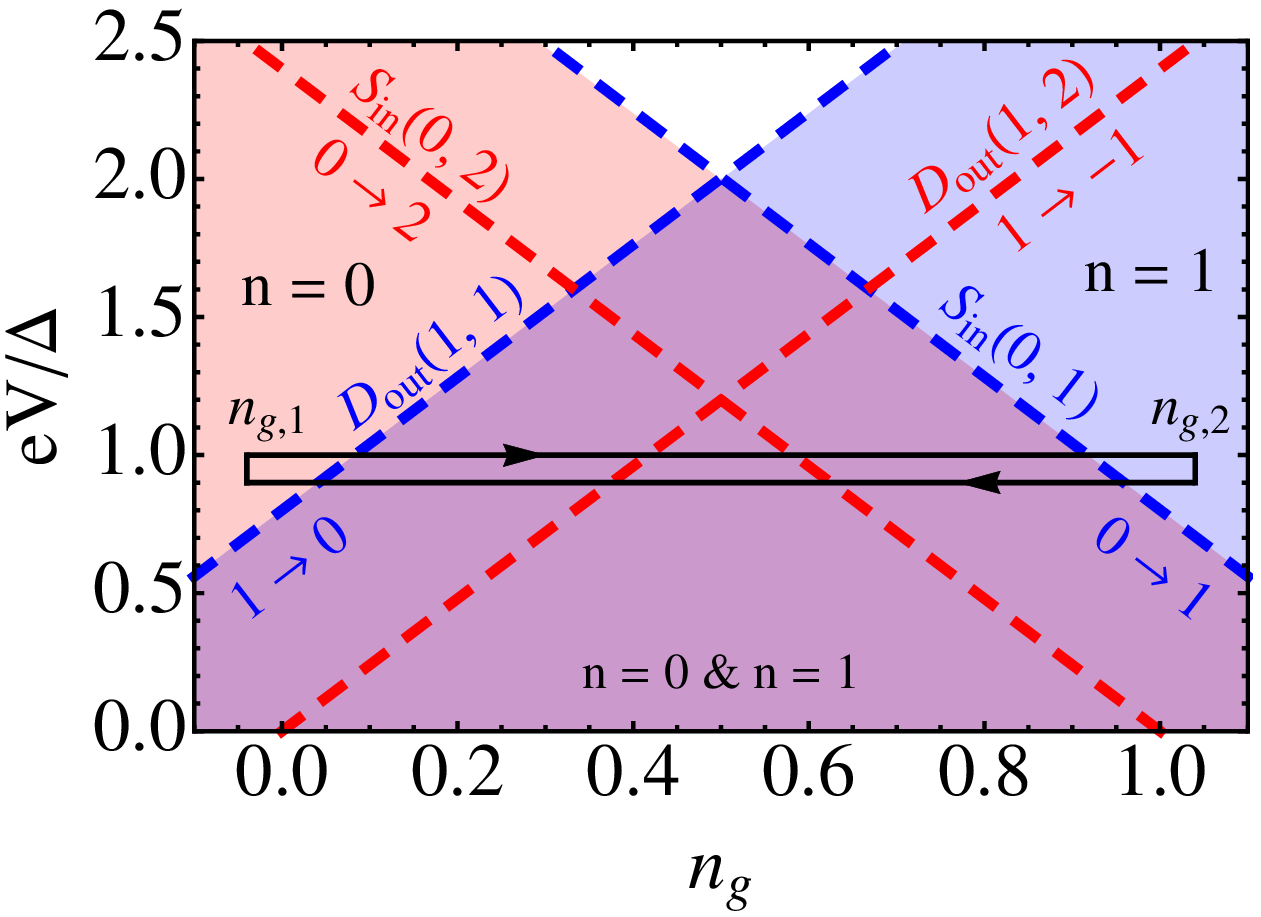}\label{fig:diamonds_andreev1}}
\hspace{3mm} \subfigure[]
{\includegraphics[height=3.9cm]{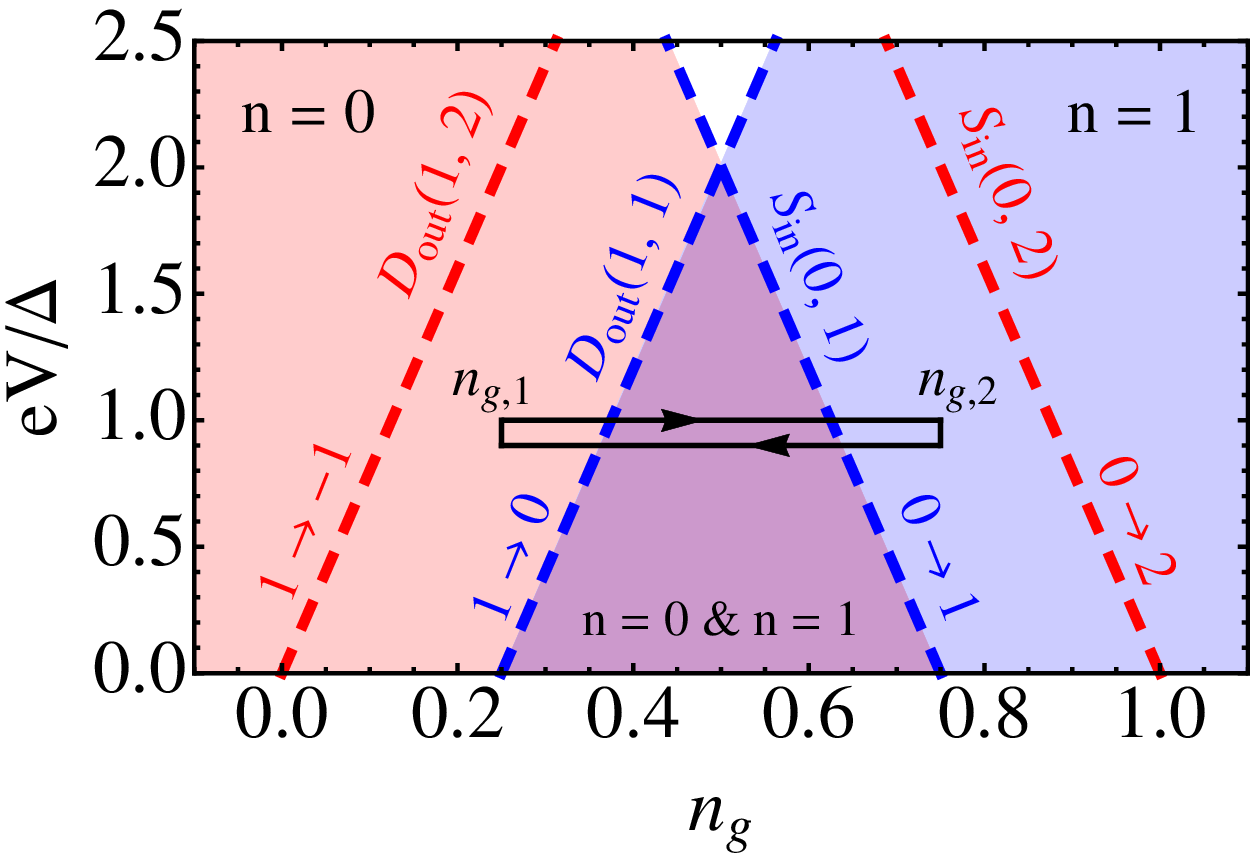}\label{fig:diamonds_andreev2}}
\caption{(Color online) Close view of the overlapping region between the Coulomb diamonds for $n=0$ and $n=1$ obtained using (a) $E_C/\Delta = 1$, (b) $E_C/\Delta = 0.6$, and (c) $E_C / \Delta =2$. Also shown are the single- (dashed blue lines) and two-particle (dashed red lines) thresholds and the optimal loop (solid black lines) at $eV \simeq \Delta$ from $n_g=n_{g,1}$ to  $n_g=n_{g,2}$.}
\end{figure*}

In the Coulomb blockade regime, the electronic transport in the SINIS device of Fig.~\ref{fig:circuit} is determined by the charging energy $E_C$. For a symmetric device, $C_S=C_D=C$, assuming that initially the excess electric charge localized on the island is $-ne$, with $n$ an integer, the energy cost to add ($+N$, in) or remove ($-N$, out) $N$ extra-electrons to or from the central normal-metal electrode is given by
\begin{eqnarray}
\label{eq:E_D}
E_D^\textrm{in/out}(n,N) & \equiv & E_\textrm{island}^D(n \pm N)-E_\textrm{island}^D(n)= \nonumber \\
&=& E_C N^2\pm\frac{1}{2}eVN\pm 2E_C (n-n_g)N \ ,
\end{eqnarray}
if the tunneling process occurs through the drain (D), and
\begin{eqnarray}
\label{eq:E_S}
E_S^\textrm{in/out}(n,N) & \equiv & E_\textrm{island}^S(n \pm N)-E_\textrm{island}^S(n)= \nonumber \\
&=& E_C N^2\mp\frac{1}{2}eVN\pm 2E_C (n-n_g)N \ ,
\end{eqnarray}
when the insulating barrier of the source (S) is overcome. In Eqs.~(\ref{eq:E_D}) and (\ref{eq:E_S}), the total energy of the island $E_\textrm{island}^{i}(n \pm N)$, with $i=S,D$, is the difference between the electrostatic energy due to the Coulomb interactions involving also the induced charge, and the work done by all the voltage sources to increase or decrease $n$ with the tunneling of $N$ particles through one of the insulating layers; $n_g=C_g V_g /e$ is the gate-induced charge.~\cite{Tinkham:1996,Ingold:1992}

\subsection{Single-Electron Tunneling}

Due to the energy gap in the BCS density of states of a superconductor, single-electron tunneling events ($N=1$) are energetically allowed above the gap, i.e., when the changes in energy Eqs.~(\ref{eq:E_D}) and (\ref{eq:E_S}) are smaller than $-\Delta$. On the contrary, above $-\Delta$ the excess charge $-ne$ of the island remains fixed to its initial value. Per each $n$, the threshold conditions $E_D^\textrm{in/out}(n,1) = -\Delta$ and $E_S^\textrm{in/out}(n,1) = -\Delta$ give rise to four crossing lines in the plot of the total bias voltage $V$ as a function of the gate-induced charge $n_g$. The four intersection points between these lines are the edges of the so-called Coulomb diamond which is a stability region for the system. This means that no single-electron tunneling process can occur for the values of $V$ and $n_g$ within its area. Unlike the case of a fully normal SET, NININ, the Coulomb diamonds for a SINIS device corresponding to different $n$ overlap. Specifically, when $E_C \sim \Delta$, the stability region for a given $n$ shares two distinct portions of the $V$ vs $n_g$ plane with the $n + 1$ and $n - 1$ diamonds, i.e., in each overlapping area at most two different values of $n$ are stable. This feature is at the basis of the generation of a controlled and synchronized single-electron current through the hybrid single-island structure of Fig.~\ref{fig:circuit}. In this regard, let us consider, for instance, the plot of Fig.~\ref{fig:diamonds_single} where a close view of the Coulomb diamonds corresponding to $n=0$ and $n=1$ and their shared part are shown. In principle, to have a cycle corresponding to a single-particle transfer from the source to the drain, $n_g$ has to move along a closed path in the $V$ vs $n_g$ plane which connects the diamonds where $n=0$ and $n=1$ are stable. Thanks to the presence of the overlapping region, this kind of connection can be realized avoiding the part of the plane where both $n=0$ and $n=1$ are unstable. As a result, each single-electron tunneling event to/from the central island can be controlled by means of the gate potential $V_g$. During each cycle of $n_g$ along the working loop, the bias voltage $V$ is usually kept fixed close to $\Delta /e$. For this optimal value, the superconducting energy gap $\Delta$ guarantees an efficient suppression of thermally-activated tunneling events and quasi-particle excitations as well as elastic and inelastic cotunneling processes.~\cite{Averin:2008,Pekola:2008} A typical loop used in real experiments with these features is shown in Fig.~\ref{fig:diamonds_single}. Starting from $n_g=n_{g,1}$, the number of excess electrons localized on island, whose initial value is $n=0$, remains constant until the threshold $S_\textrm{in}(0,1)$, defined by the equation $E_S^\textrm{in}(0,1)=-\Delta$, is crossed. At that point one electron can enter in the central electrode via the source junction and $n$ passes from $0$ to $1$. Once $n_g=n_{g,2}$ is reached, the closed path is covered backward. The extra electron on the island can tunnel out through the drain only after overcoming the threshold $D_\textrm{out}(1,1)$, given by the equation $E_D^\textrm{out}(1,1)=-\Delta$. When $n_g$ is again equal to $n_{g,1}$, the island is back in its initial state and a new cycle can start. Since per each cycle exactly one electron is transferred from the source to the drain, driving $n_g$ from $n_{g,1}$ to $n_{g,2}$ and back to $n_{g,1}$ with a signal with frequency $f$ allows to generate the single-electron current $ I = e f $.

\subsection{Andreev Reflection and Higher-Order Processes}

In addition to single-electron tunneling events, the current flowing through a SINIS transistor is, in general, also affected by the Andreev reflection, i.e., the transfer of two-electrons per unit of time inside or outside the island.~\cite{Andreev:1964,HekkingGlazman:1993} This second-order tunneling process is insensitive to the energy barrier provided by the superconducting gap. This means that the rate of the transitions $n\rightarrow n\pm 2$ can be relevant although the device is working at the optimal bias, $eV \simeq \Delta$. As a result, the Coulomb diamonds for the Andreev reflection events are obtained just imposing that the energies Eqs.~(\ref{eq:E_D}) and (\ref{eq:E_S}) for $N=2$ are smaller than zero. However, as shown in Figs.~\ref{fig:diamonds_andreev1} and \ref{fig:diamonds_andreev2}, the energy $\Delta$, together with the charging energy $E_C$, plays an important role in the determination of the two-electron tunneling probability. If the ratio $E_C / \Delta$ is smaller than 1, the Andreev diamonds are contained within the single-particle stability regions. In this case, we see from Fig.~\ref{fig:diamonds_andreev1} that the optimal loop crosses the two-particle threshold $S_\textrm{in}(0,2)$, given by the equation $E_S^\textrm{in}(0,2)=0$, before the single-electron line $S_\textrm{in}(0,1)$, while going from $n_{g,1}$ to $n_{g,2}$. When $n_g$ is decreased back to $n_{g,1}$, the closed path overcomes $D_\textrm{out}(1,1)$ after $D_\textrm{out}(1,2)$, the line corresponding to $E_D^\textrm{out}(1,2)=0$. It follows that, in this regime, the control of single-electron tunneling is compromised by the Andreev transitions $0\rightarrow 2$ and $1\rightarrow -1$. On the other hand, when $E_C / \Delta >1$, the single-particle diamonds are smaller than the ones for Andreev reflection. Now, the two-particle thresholds can be avoided, as shown in Fig.~\ref{fig:diamonds_andreev2}, thereby suppressing the probability to increase/decrease the charge of the island by two electrons per each tunneling event.

However, higher-order processes, such as the cotunneling of one electron and one Cooper-pair,~\cite{Averin:2008} can occur while $n_g$ covers the loop of Fig.~\ref{fig:diamonds_andreev2}. They can limit the single-electron transfer accuracy even if $E_C / \Delta >1$. In particular, the more the system stays in the overlapping region where more than one charge state is stable, the bigger the effect of unwanted transitions would be. To decrease the influence of the higher-order error events, the signal $n_g(t)$ which is usually used to go from $n_{g,1}$ to $n_{g,2}$ and back to $n_{g,1}$ is a square-wave. This choice guarantees that the time spent in between $n_{g,1}$ and $n_{g,2}$ is minimized. On the other hand, the period $\tau=1/f$ of $n_g(t)$ has to be long enough in order for the single-particle tunneling processes to take place. If the number $n$ changes by one electron with the rate $\Gamma_{1 e}$, then the tunneling error or probability that the charge of the island remains the same is $\varepsilon \sim \exp(-\Gamma_{1e} /2f )$. In particular, the requirement $\varepsilon  \lesssim \varepsilon_\textrm{metr} = 10^{-8}$ has to be satisfied for the definition of the quantum current standard. This means that $\Gamma_{1e} \gtrsim 10^9\unit{s^{-1}}$ because the trade-off between the missed tunneling discussed above and the leakage by Cooper pair - electron cotunneling limits the maximum operation frequency to $f \sim 60\unit{MHz}$ to achieve the metrologically-accurate current $I=ef \sim 10\unit{pA}$ for a single SINIS turnstile.~\cite{Averin:2008,Kemppinen:2009,Kemppinen:2009.2}

\section{Environment-assisted Andreev reflection}

\subsection{The effect of the electromagnetic environment on the electronic transport}

As discussed in the previous section, the tunneling processes involving more than one electron may be reduced biasing the SINIS turnstile at the optimal value $eV\simeq \Delta$, considering $E_C/\Delta > 1$ and using for $n_g(t)$ a square-wave-like signal which oscillates with frequency $f$ between the two induced gate charges $n_{g,1}$ and $n_{g,2}$ of Fig.~\ref{fig:diamonds_andreev2}. Under these conditions, one expects to measure the current $I=ef$ with a relatively high accuracy. In principle, it should be possible even going below the relative error $\varepsilon_\textrm{metr}$ required by the metrological applications. However, in real experiments, the achievement of the accuracy needed for the definition of the quantum current standard still remains a difficult task.

\begin{figure}[ht!]
\includegraphics[height=3.5cm]{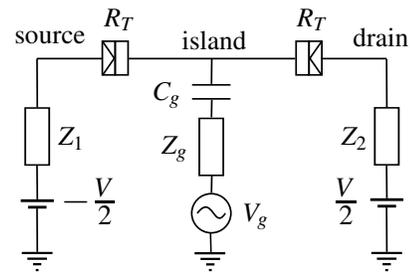}
\caption{Circuit representation of the hybrid S-I-N-I-S single-electron transistor (SET). The two NIS junctions constituting the device have the same capacitance $C$ and tunnel resistance $R_T$ and are connected to the source $V_S=-V/2$ and drain $V_D=V/2$ voltages via the impedances $Z_1(\omega)$ and $Z_2(\omega)$ respectively. The normal metal island is controlled by means of the gate voltage $V_g$ via the capacitance $C_g$. The gate impedance $Z_g(\omega)$ together with $Z_1(\omega)$ and $Z_2(\omega)$ represent the electromagnetic environment at temperature $T_\textrm{env}$.}
\label{fig:circuitBIS}
\end{figure}

The coupling of the hybrid turnstile with its surrounding high-temperature electromagnetic environment may be a detrimental source of error.~\cite{Pekola:2010} Indeed, the absorption/emission of energy from/to such a thermal bath allows the tunneling of electrons, even when the overcoming of the insulating barrier results to be energetically forbidden for a well isolated SET. Nevertheless, the environment-assisted tunneling of quasi-particles can be efficiently suppressed using, for instance, an on chip capacitively coupled ground plane~\cite{Pekola:2010} and/or by means of a highly-resistive transmission line.~\cite{DiMarco} The main contribution to the leakage current observed in the I-V characteristic is typically due to the Andreev reflection. Although large charging energies, $E_C>\Delta$, should reduce the probability for this two-particle process to occur, the tunneling of Cooper-pairs still can have a strong influence on the current flowing through the transistor.~\cite{Kemppinen:2009,Aref:2011} The enhancement of the Andreev tunneling events due to the coupling of the system with the external bath may account for this behavior.\cite{Bubanja:2014} To understand under which conditions the environment-assisted Andreev reflection can be relevant, we consider the circuit of Fig.~\ref{fig:circuitBIS} where we introduce the effective impedances $Z_1(\omega)$, $Z_2(\omega)$, and $Z_g(\omega)$ to model the thermal bath. We assume also that the two junctions in the system have the same tunnel resistance $R_T$.

\subsection{Single-photon-assisted two-electron tunneling rate}

In order to find the tunneling rate of the Andreev reflection process under the effect of the electromagnetic environment, we start by considering the tunneling Hamiltonian
\begin{equation}
\label{eq:H_tunnel}
\hat{H}_T = e^{i \hat{\varphi}_\textrm{env}} \sum_{\mathbf{k},\mathbf{p},\sigma} t_{\mathbf{k},\mathbf{p}} \left( u_\mathbf{p} \hat{\gamma}^\dag_{\mathbf{p},\sigma} + v_\mathbf{p} \hat{\gamma}_{-\mathbf{p},-\sigma}  \right) \hat{a}_{\mathbf{k},\sigma} + \textrm{H.c.}  \  ,
\end{equation}
which accounts for the transfer of two electrons between the normal-metal island and one of the superconducting electrodes of the SINIS SET of Fig.~\ref{fig:circuitBIS}. Equation (\ref{eq:H_tunnel}) is written in terms of the creation $\hat{\gamma}_{\mathbf{p},\sigma}^\dag$ ($\hat{a}^\dag_{\mathbf{k},\sigma}$) and annihilation $\hat{\gamma}_{\mathbf{p},\sigma}$ ($\hat{a}_{\mathbf{k},\sigma}$) operators of quasiparticles (electrons) in S (N) with wave vector $\mathbf{p}$ ($\mathbf{k}$) and spin $\sigma= \uparrow , \downarrow$. The tunnel matrix elements $t_{\mathbf{k},\mathbf{p}}$, in general, depend on $\mathbf{p}$ and $\mathbf{k}$. The BCS coherence factors $u_\mathbf{p}$ and $v_\mathbf{p}$ are spin-independent and satisfy the relations
\begin{equation}
\label{eq:BCS_coherence}
u_\mathbf{p}^2 = 1- v_\mathbf{p}^2 = \frac{1}{2}\left( 1 + \frac{\xi_\mathbf{p}}{\varepsilon_\mathbf{p}} \right)  \quad , \quad  u_\mathbf{p} v_\mathbf{p} = \frac{\Delta}{(\varepsilon_\mathbf{p}^2 - \Delta^2)^{1/2}} \ ,
\end{equation}
where $\xi_\mathbf{p}$ is the energy of an electron in S with momentum $\mathbf{p}$ measured with respect to the Fermi level, and $\varepsilon_\mathbf{p} = (\xi_\mathbf{p}^2 + \Delta^2)^{1/2}$ is the quasiparticle energy. The translation operator $e^{i \hat{\varphi}_\textrm{env}}$ in Eq.~(\ref{eq:H_tunnel}) accounts for the change of the charge of the electrodes due to the environment-assisted tunneling of one electron. Considering the environment as an infinite ensemble of quantum harmonic oscillators with temperature $T_\textrm{env}$ (Caldeira-Leggett model~\cite{Caldeira:1983,Caldeira:1983erratum,Leggett:1987}), the fluctuating phase $\hat{\varphi}_\textrm{env}$ can be written as
\begin{equation}
\label{eq:total_phase}
\hat{\varphi}_\textrm{env} = \sum_\lambda \hat{\varphi}_\lambda = \sum_\lambda \rho_\lambda \left(  \hat{c}_\lambda^\dag +  \hat{c}_\lambda \right)
\end{equation}
where the phase $\hat{\varphi}_\lambda$ represents the position operator of the harmonic oscillator $\lambda$ with mass $C_\lambda$ and characteristic frequency $\omega_\lambda= 1/ \sqrt{L_\lambda C_\lambda}$. The coupling term is $\rho_\lambda=(e/\hbar)\sqrt{\hbar / 2 C_\lambda \omega_\lambda}$, and the operators $\hat{c}_\lambda^\dag$ and  $\hat{c}_\lambda$ create and annihilate one photon with energy $\hbar \omega_\lambda$ (see Appendix~\ref{app:Caldeira_FluctDiss}). Hereafter, we assume that the coupling of the SINIS with the environment is weak, meaning that at most a single photon is involved in the exchange of energy between the system and the thermal bath.~\cite{DiMarco} In other words, we consider the limit where $\rho_\lambda \ll 1$ and the series expansion of the charge translation operator Eq.~(\ref{eq:total_phase}) in $\hat{H}_T$ can be truncated at the first order, i.e., $e^{i \hat{\varphi}_\textrm{env}} \simeq 1 + i \hat{\varphi}_\textrm{env}$. The validity of this assumption will be discussed in the following.

\begin{figure}[t!]
\includegraphics[height=2.6cm]{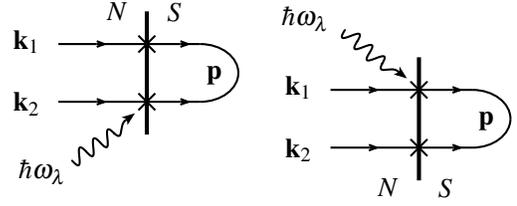}
\caption{Feynman diagrams of the two possible single-photon absorption processes giving rise to the environment-assisted Andreev reflection.}
\label{fig:Feynman}
\end{figure}

Let us focus on the Andreev process $1\rightarrow -1$, characterized by the transfer of two electrons from the normal metal island to the superconducting drain electrode as a Cooper pair. According to perturbation theory in $\hat{H}_T$, the total probability amplitude to have such a second-order event in the system of Fig.~\ref{fig:circuitBIS} is given by
\begin{equation}
\label{eq:Andreev_ampl1}
A_{\mathbf{k}_1,\mathbf{k}_2}^\lambda=\sum_{m_\lambda} \frac{\left\langle f_\lambda | \hat{H}_T | m_\lambda \right\rangle  \left\langle m_\lambda | \hat{H}_T | i_\lambda \right\rangle }{\zeta_{m_\lambda} - \zeta_{i_\lambda} + i\eta}  \ ,
\end{equation}
for fixed values of the environmental index $\lambda$, and of the initial wave vectors $\mathbf{k}_1$ and $\mathbf{k}_2$. Here the initial state is
\begin{equation}
\label{eq:In_state}
| i_\lambda\rangle  =   | \mathbf{k}_1 \uparrow,\mathbf{k}_2 \downarrow \rangle_N \otimes | n_\textrm{pairs}, \xcancel{\mathbf{p}} \rangle_S \otimes | n_\lambda +1 \rangle_\textrm{env} \ ,
\end{equation}
with two electrons in N with opposite spin and momenta $\mathbf{k}_1$ and $\mathbf{k}_2$, $ n_\textrm{pairs}$ Cooper pairs and no quasiparticle excitations in S, and $n_\lambda + 1$ photons with energy $\hbar \omega_\lambda$ in the environment. On the other hand, the final state is
\begin{equation}
\label{eq:Fin_state}
| f_\lambda \rangle  =  | \xcancel{\mathbf{k}_1 \uparrow },\xcancel{\mathbf{k}_2 \downarrow}\rangle_N \otimes | n_\textrm{pairs} + 1, \xcancel{\mathbf{p}}  \rangle_S \otimes | n_\lambda \rangle_\textrm{env}  \  ,
\end{equation}
with an additional Cooper pair in S, two less electrons in N, and one less photon in the Caldeira-Leggett bath. The transition from the state (\ref{eq:In_state}) to the state (\ref{eq:Fin_state}) is determined by all the possible intermediate virtual states $ | m_\lambda \rangle$ such that a quasi-particle with momentum $\mathbf{p}$ is created in S after the annihilation of one of the two electrons in N. As illustrated in Fig.~(\ref{fig:Feynman}), only one of the two tunneling electrons can absorb the energy of the only available photon, in the weak coupling limit. As a result, for a fixed wave vector $\mathbf{p}$ of the virtual quasiparticle in S, only the four intermediate states
\begin{eqnarray}
\label{eq:Inter_states}
| {1_\lambda} \rangle & = & |\mathbf{k}_1 \uparrow ,\xcancel{\mathbf{k}_2 \downarrow }\rangle_N \otimes | n_\textrm{pairs}, \mathbf{p} \rangle_S \otimes | n_\lambda +1 \rangle_\textrm{env} \ , \nonumber \\
| {2_\lambda} \rangle & = & |\mathbf{k}_1 \uparrow ,\xcancel{\mathbf{k}_2 \downarrow }\rangle_N \otimes | n_\textrm{pairs}, \mathbf{p} \rangle_S \otimes | n_\lambda  \rangle_\textrm{env} \ , \nonumber \\
| {3_\lambda} \rangle & = & | \xcancel{\mathbf{k}_1 \uparrow },\mathbf{k}_2 \downarrow \rangle_N \otimes | n_\textrm{pairs}, \mathbf{p} \rangle_S \otimes | n_\lambda +1 \rangle_\textrm{env} \ , \nonumber \\
| {4_\lambda} \rangle & = & | \xcancel{\mathbf{k}_1 \uparrow },\mathbf{k}_2 \downarrow \rangle_N \otimes | n_\textrm{pairs}, \mathbf{p} \rangle_S \otimes | n_\lambda  \rangle_\textrm{env} \ ,
\end{eqnarray}
can give a non-zero contribution to $A_{\mathbf{k}_1,\mathbf{k}_2}^\lambda$. The difference between the energies $\zeta_{m_\lambda}$ of these virtual states and the energy $\zeta_{i_\lambda}-i\eta$ of the initial state $| i_\lambda\rangle$ determine the amplitude Eq.~(\ref{eq:Andreev_ampl1}). The imaginary part  $\eta = \hbar \Gamma_{1 \rightarrow 0} /2$ accounts for the lifetime broadening of $| i_\lambda\rangle$ due to the competing single-electron tunneling processes occurring with rate $\Gamma_{1 \rightarrow 0}$. According to perturbation theory in the tunneling Hamiltonian $\hat{H}_T$, the first-order rate, describing one electron going out of the island through the drain, can be written as
\begin{equation}
\label{eq:single_rate}
\Gamma_{1 \rightarrow 0}^\textrm{\tiny{Dynes}} = \frac{1}{2\pi}\frac{\Delta}{\hbar} \frac{R_K}{R_T} \int_0^{\left| E_D^\textrm{out}(1,1) \right|} \frac{N_S^\textrm{\tiny{Dynes}}(E/\Delta)}{\Delta} \  dE
\end{equation}
in terms of the Dynes density of states of a superconductor,~\cite{Dynes:1978}
\begin{equation}
N_S^\textrm{\tiny{Dynes}}(E/\Delta)=\left|  \Re\mbox{e}\left[ \frac{E/\Delta + i \gamma_\textrm{\tiny{Dynes}}}{\sqrt{(E/\Delta + i \gamma_\textrm{\tiny{Dynes}})^2-1}}  \right]  \right| \ ,
\end{equation} 
which depends on the phenomenological Dynes parameter $\gamma_\textrm{\tiny{Dynes}}$. In Eq.~(\ref{eq:single_rate}), $E_D^\textrm{out}(1,1) = 2E_C(n_g - 1/2) - eV/2$ is the energy cost that has to be payed by the voltage sources in order for the transition $1 \rightarrow 0$ to occur [see Eq.~(\ref{eq:E_D})]; $R_K=h/e^2$ is the resistance quantum.
The Dynes rate Eq.~(\ref{eq:single_rate}) is valid in the zero-temperature limit, $k_B T_\textrm{SINIS} \ll \Delta$, and takes into account the most relevant single-electron error sources, such as the environment.

\begin{figure*}[ht!]
\centering  \subfigure[]
{\includegraphics[height=5.2cm]{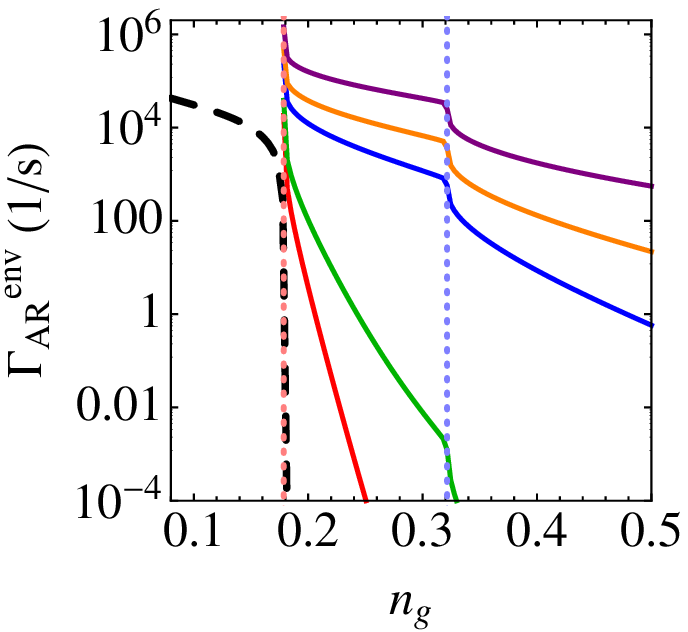}\label{fig:Rates_diff_T}}
\hspace{1mm} \subfigure[]
{\includegraphics[height=5.2cm]{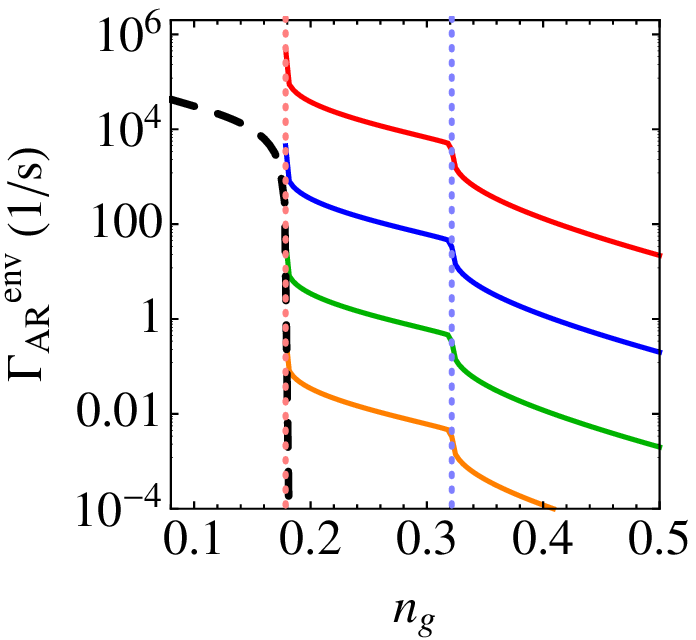}\label{fig:Rates_diff_R}}
\hspace{1mm} \subfigure[]
{\includegraphics[height=5.2cm]{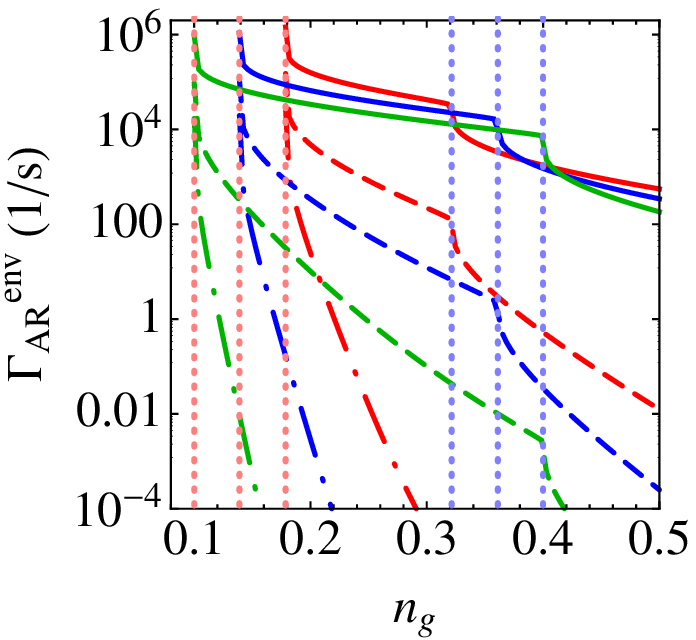}\label{fig:Rates_diff_Ec}}
\caption{(Color online) Photon-assisted Andreev rates, given by the numerical evaluation of Eq.~(\ref{eq:Rate_final}), as a function of the gate-induced charge $n_g$ with $\Delta=210\unit{\mu eV}$ (Aluminum), $R_T=430\unit{k\Omega}$, $C_g=10^{-16}$, $\mathcal{N}=100$, and $\gamma_\textrm{\tiny{Dynes}}=10^{-5}$. In panel (a), for each rate $R=1100\unit{\Omega}$ and $E_C/\Delta = 1.4$ with $C=0.86 \cdot 10^{-16}$; the values of $T_\textrm{env}$ are: $70\unit{mK}$ (red), $140\unit{mK}$ (blue), $780\unit{mK}$ (green), $1.5\unit{K}$ (orange), $4.2\unit{K}$ (purple). In panel (b), for each curve $T_\textrm{env}=1.5\unit{K}$ and $E_C/\Delta = 1.4$ with $C=0.86 \cdot 10^{-16}$; the resistances $R$ are: $1100\unit{\Omega}$ (red), $10\unit{\Omega}$ (blue), $0.1\unit{\Omega}$ (green), $0.001\unit{\Omega}$ (orange). In both panel (a) and (b), the dashed black line is the Andreev rate valid in the absence of environment (see Ref.~\onlinecite{Averin:2008}). In panel (c), for fixed $R=1100\unit{\Omega}$, the curves with the same color are obtained using the same charging energy, $E_C/\Delta$: 1.4 with $C=0.86 \cdot 10^{-16}$ (red lines), 1.8  with $C=0.558 \cdot 10^{-16}$ (blue lines), 2.5  with $C=0.262 \cdot 10^{-16}$ (green lines); the values of $T_\textrm{env}$ are: $4.2\unit{K}$ (solid curves), $500\unit{mK}$ (dashed curves), and $100\unit{mK}$ (dotted-dashed curves). In all the three panels, also shown are the single- and two-particle thresholds, $1/2-\Delta/4E_C$ (light-blue vertical dotted lines), and $\Delta/4E_C$ (light-red vertical dotted lines), respectively.}
\end{figure*}

\begin{figure*}[ht!]
\centering  \subfigure[]
{\includegraphics[height=5.2cm]{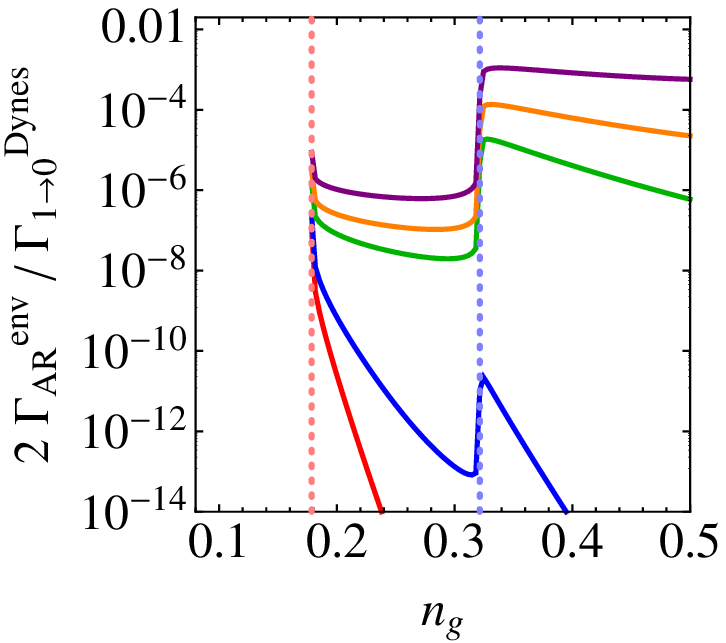}\label{fig:Ratios_diff_T}}
\hspace{8mm} \subfigure[]
{\includegraphics[height=5.2cm]{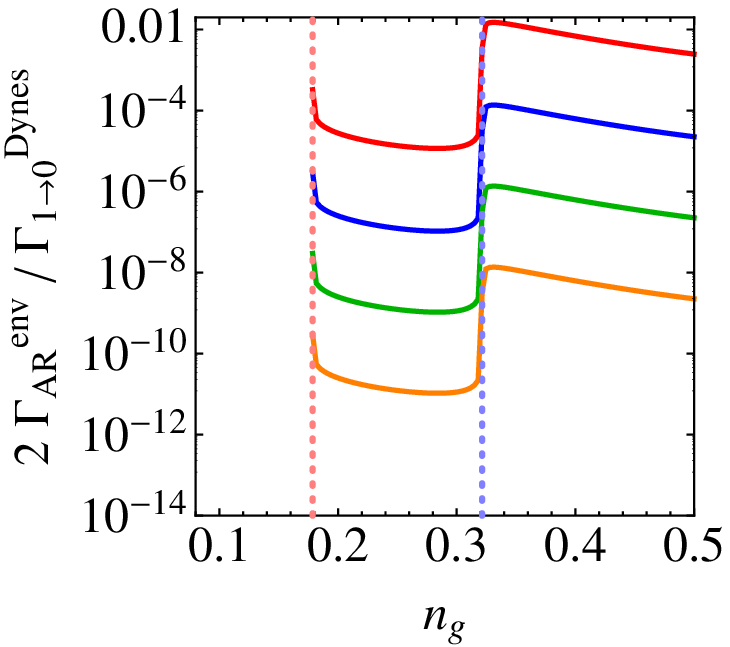}\label{fig:Ratios_diff_R}}
\hspace{1mm} \subfigure[]
{\includegraphics[height=5.2cm]{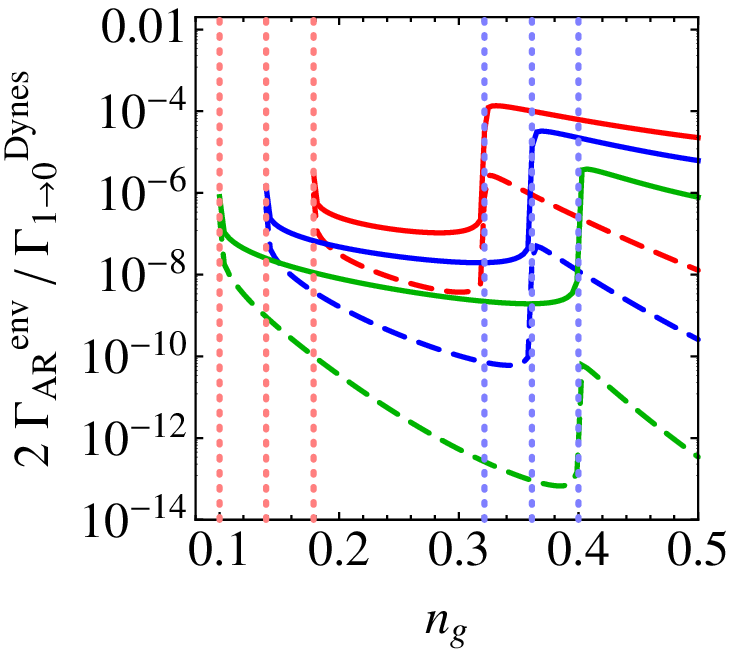}\label{fig:Ratios_diff_Ec}}
\hspace{8mm} \subfigure[]
{\includegraphics[height=5.2cm]{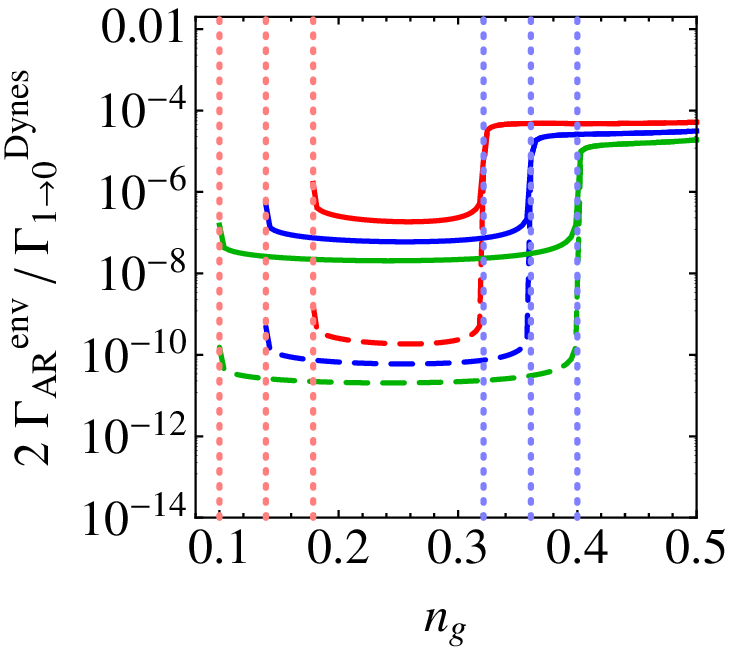}\label{fig:RatiosHighT}}
\caption{(Color online) Plot of the ratio $\varepsilon_\textrm{acc}$ as a function of the gate-induced charge $n_g$ with $\Delta=210\unit{\mu eV}$ (Aluminum), $R_T=430\unit{k\Omega}$, $C_g=10^{-16}$, and $\mathcal{N}=100$. In panel (a), for each rate $\gamma_\textrm{\tiny{Dynes}}=10^{-5}$, $R=10\unit{\Omega}$ and $E_C/\Delta = 1.4$ with  $C=0.86 \cdot 10^{-16}$; the values of $T_\textrm{env}$ are: $70\unit{mK}$ (red), $140\unit{mK}$ (blue), $780\unit{mK}$ (green), $1.5\unit{K}$ (orange), $4.2\unit{K}$ (purple). In panel (b), for each curve $\gamma_\textrm{\tiny{Dynes}}=10^{-5}$, $T_\textrm{env}=1.5\unit{K}$ and $E_C/\Delta = 1.4$ with  $C=0.86 \cdot 10^{-16}$; the resistances $R$ are: $1100\unit{\Omega}$ (red), $10\unit{\Omega}$ (blue), $0.1\unit{\Omega}$ (green), $0.001\unit{\Omega}$ (orange). In panel (c), for fixed $\gamma_\textrm{\tiny{Dynes}}=10^{-5}$ and $R=10\unit{\Omega}$, the curves with the same color are obtained using the same charging energy, $E_C/\Delta$: 1.4 with  $C=0.86 \cdot 10^{-16}$ (red lines), 1.8 with  $C=0.558 \cdot 10^{-16}$ (blue lines), 2.5 with  $C=0.262 \cdot 10^{-16}$ (green lines); the values of $T_\textrm{env}$ are: $1.5\unit{K}$ (solid curves), and $500\unit{mK}$ (dashed curves). In panel (d), the Dynes parameter $\gamma_\textrm{\tiny{Dynes}}$ is equal to $10^{-4}$ (solid lines) and  $10^{-7}$ (dashed lines). The curves with the same color are obtained using the same charging energy, $E_C/\Delta$: 1.4 (red lines), 1.8 (blue lines), 2.5 (green lines). In all the four panels, the single- and two-particle thresholds, $1/2-\Delta/4E_C$ (light blue vertical dotted lines), and $\Delta/4E_C$ (light red vertical dotted lines) respectively are also shown.}\label{fig:Ratios}
\end{figure*}

Using Eqs.~(\ref{eq:In_state}), (\ref{eq:Fin_state}), and (\ref{eq:Inter_states}), the amplitude Eq.~(\ref{eq:Andreev_ampl1}) reads as
\begin{equation}
\label{eq:Andreev_ampl2}
A_{\mathbf{k}_1,\mathbf{k}_2}^\lambda  = i \ t_0^2 \ \sqrt{f_{\mathbf{k}_1}} \sqrt{f_{\mathbf{k}_2}} \ \rho_\lambda \sqrt{n_\lambda} \ \sum_{\mathbf{p}} \ \big( u_\mathbf{p} v_\mathbf{p} \big) \ S_{\mathbf{p},\lambda} \ ,
\end{equation}
for a low-temperature hybrid single-electron transistor, $k_B T_\textrm{SINIS} \ll \Delta$, and assuming constant tunneling matrix elements, $t_{\mathbf{k},\mathbf{p}}=t^*_{\mathbf{k},\mathbf{p}}=t_0$ (point tunnel junction). In Eq.~(\ref{eq:Andreev_ampl2}), we introduced the Fermi-Dirac distribution function $f_{\mathbf{k}}=[\exp(\xi_\mathbf{k}/k_B T_\textrm{SINIS})+1]^{-1}$ for the normal metal electrons and the sum of the intermediate-state denominators
\begin{eqnarray}
\label{eq:Sum}
S_{\mathbf{p},\lambda} & \equiv & \ \frac{1}{\varepsilon_{\mathbf{p}}^\textrm{c}-\xi_{\mathbf{k}_1}+i\eta} \ + \ \frac{1}{\varepsilon_{\mathbf{p}}^\textrm{c}-\xi_{\mathbf{k}_2}-\hbar\omega_\lambda+i\eta}  \nonumber  \\
& + &  \ \frac{1}{\varepsilon_{\mathbf{p}}^\textrm{c}-\xi_{\mathbf{k}_2}+i\eta} \ + \ \frac{1}{\varepsilon_{\mathbf{p}}^\textrm{c}-\xi_{\mathbf{k}_1}-\hbar\omega_\lambda+i\eta} \ .
\end{eqnarray}
Here $\varepsilon_{\mathbf{p}}^\textrm{c} \equiv \varepsilon_{\mathbf{p}}+E_D^\textrm{out}(1,1)$ is the virtual state energy and $\xi_{\mathbf{k}}$ is the energy of an electron in N with momentum $\mathbf{k}$ measured with respect to the Fermi level. Summing over all the possible initial states and considering the spin degeneracy, one obtains the total rate 
\begin{equation}
\label{eq:Rate}
\Gamma_\textrm{AR}^\textrm{ env} = \frac{4\pi}{\hbar} \sum_{\mathbf{k}_1,\mathbf{k}_2} \sum_\lambda \left| A_{\mathbf{k}_1,\mathbf{k}_2}^\lambda \right|^2  \delta\left(  \xi_{\mathbf{k}_1,\mathbf{k}_2}^\textrm{c} + \hbar\omega_\lambda  \right) \ ,
\end{equation}
where $\xi_{\mathbf{k}_1,\mathbf{k}_2}^\textrm{c} \equiv \xi_{\mathbf{k}_1} + \xi_{\mathbf{k}_2} - E_D^\textrm{out}(1,2)$ is determined by the energy cost $E_D^\textrm{out}(1,2) = 4E_C n_g - eV$ needed for the second-order transition $1 \rightarrow -1$ to occur [see Eq.~(\ref{eq:E_D})]. The environment-assisted Andreev rate Eq.~(\ref{eq:Rate}) is written in terms of the probability
$$
\left| A_{\mathbf{k}_1,\mathbf{k}_2}^\lambda \right|^2 = t_0^4 \ f_{\mathbf{k}_1} f_{\mathbf{k}_2} \ \rho_\lambda^2 \ n_\lambda \ \sum_{\mathbf{p},\mathbf{p'}} \big( u_\mathbf{p} v_\mathbf{p} \big) \big( u_\mathbf{p'} v_\mathbf{p'} \big)  \ S_{\mathbf{p},\lambda} \ S_{\mathbf{p'},\lambda}^*  \  .
$$
Approximating the sums over $\mathbf{k}_1$, $\mathbf{k}_2$,  $\mathbf{p}$ and $\mathbf{p'}$ by the corresponding integrals, assuming that $n_\lambda$ is given by the Bose-Einstein distribution, $n_{BE}(\omega_\lambda) = [\exp(\hbar\omega_\lambda/$ $k_BT_\textrm{env})-1]^{-1}$, and using the properties of the Dirac delta function, Eq.~(\ref{eq:Rate}) can be written as
\begin{eqnarray}
\label{eq:Rate2}
\Gamma_\textrm{AR}^\textrm{ env} & \simeq & \frac{1}{(2 \pi)^3}\frac{1}{2 \hbar}\left( \frac{R_K}{R_T} \right)^{\!\!\! 2} \int_{-\infty}^{+\infty} d\xi_{\mathbf{k}_1} \ d\xi_{\mathbf{k}_2} \ \int_{-\infty}^{+\infty} d\xi_{\mathbf{p}} \ d\xi_{\mathbf{p'}} \nonumber \\
& \times & \big( u_\mathbf{p} v_\mathbf{p} \big) \ \big( u_\mathbf{p'} v_\mathbf{p'} \big) \ f_{\mathbf{k}_1} f_{\mathbf{k}_2} \ n_{BE}\left( -\xi_{\mathbf{k}_1,\mathbf{k}_2}^\textrm{c} \right) / \left( -\xi_{\mathbf{k}_1,\mathbf{k}_2}^\textrm{c} \right)  \nonumber \\
& \times & \left( S_{\mathbf{p}} \ S_{\mathbf{p'}}^* \right) \ \sum_\lambda \rho_\lambda^2 \omega_\lambda \delta\left( \xi_{\mathbf{k}_1,\mathbf{k}_2}^\textrm{c} + \hbar\omega_\lambda \right) \ .
\end{eqnarray}
Here $S_{\mathbf{p}}$ and $S_{\mathbf{p'}}^*$ are $S_{\mathbf{p},\lambda}$ and $S_{\mathbf{p'},\lambda}^*$ evaluated for  $\hbar\omega_\lambda=-\xi_{\mathbf{k}_1,\mathbf{k}_2}^\textrm{c}$.  Thanks to fluctuation-dissipation theorem, the sum over $\lambda$ in Eq.~(\ref{eq:Rate2}) can be related to the effective impedance $\Re\mbox{e}\left[Z_\textrm{eff}(\omega)\right]$ ``seen'' by the drain (see Appendices \ref{app:Caldeira_FluctDiss}, \ref{app:Noise_Drain}, and \ref{app:Noise_Drain_TL}). Then, in the low-temperature limit $k_B T_\textrm{SINIS} \ll \Delta$, Eq.~(\ref{eq:Rate2}) can be recast in the form
\begin{eqnarray}
\label{eq:Rate_final}
\Gamma_\textrm{AR}^\textrm{ env} & \simeq & \frac{1}{(2 \pi)^3}\frac{\Delta^2}{\hbar R_K \mathcal{N}}\left( \frac{R_K}{R_T} \right)^{\!\!\! 2} \int_{-\infty}^{0} d\xi_{\mathbf{k}_1} \ d\xi_{\mathbf{k}_2} \int_\Delta^{+\infty} d\varepsilon_{\mathbf{p}} \ d\varepsilon_{\mathbf{p'}} \nonumber \\
& \times & \left(  \sqrt{\varepsilon_{\mathbf{p}}^2-\Delta^2} \ \sqrt{\varepsilon_{\mathbf{p'}}^2-\Delta^2}\right)^{-1}  n_{BE}\left( -\xi_{\mathbf{k}_1,\mathbf{k}_2}^\textrm{c} \right)  / \left( -\xi_{\mathbf{k}_1,\mathbf{k}_2}^\textrm{c} \right) \nonumber \\
& \times &  \left( S_{\mathbf{p}} \ S_{\mathbf{p'}}^* \right) \ \Re\mbox{e}\left[ Z_\textrm{eff} \left( \xi_{\mathbf{k}_1,\mathbf{k}_2}^\textrm{c} / \hbar  \right)\right]  \  ,
\end{eqnarray}
using the BCS relation for $u_\mathbf{p}v_\mathbf{p}$ given in Eq.~(\ref{eq:BCS_coherence}), and the quasiparticle energies $\varepsilon_{\mathbf{p}}$ and $\varepsilon_{\mathbf{p'}}$ as integration variables. In this last formula, we also introduced the number of conducting channels $\mathcal{N}$ of the junction. The environment-assisted Andreev rate Eq.~(\ref{eq:Rate_final}) is valid in the single-photon regime $\rho_\lambda \ll 1$, i.e., for small values of the effective impedance, $\Re\mbox{e}\left[Z_\textrm{eff}(\omega)\right]/R_K \ll \Delta / k_B T_\textrm{env}$.\cite{DiMarco} Additionally, since we assumed that only the absorption process can occur, one has to impose that $\xi_{\mathbf{k}_1,\mathbf{k}_2}^\textrm{c} \leq 0$ in Eq.~(\ref{eq:Rate}), namely $E_D^\textrm{out}(1,2) \geq 0$. This means that Eq.~(\ref{eq:Rate_final}) applies only for those values of the induced gate charge $n_g$ equal to and larger than the two-particle threshold $eV / 4 E_C$.

For the circuit depicted in Fig.~\ref{fig:circuitBIS}, $\Re\mbox{e}\left[Z_\textrm{eff}(\omega)\right]$ is equal to the right-hand side of Eq.~(\ref{eq:Final_Rel}). The latter is the sum of three terms which are of the same order of magnitude for the typical experimental values of the capacitances $C$ and $C_g \sim C$, and of the impedances $Z_1(\omega) \sim Z_2(\omega) \sim Z_g (\omega)$. Consequently, because of the symmetry of the circuit of Fig.~\ref{fig:circuitBIS} with respect to the gate, we focus hereafter on the case where the voltage fluctuation across $C_D$ is produced only by $Z_g(\omega)$, considering $Z_1(\omega)$ and $Z_2(\omega)$ as noiseless. In addition, we assume that the effect of $Z_1$ and $Z_2$ can be neglected with respect to $C_S$ and $C_D$, namely $Z_{1,2}[E_D^\textrm{out}(1,2)/\hbar] \ll \hbar / E_D^\textrm{out}(1,2) C_{S,D}$. As a result, setting $\delta V_1 = \delta V_2 = 0$ as well as $Z_1(\omega)=Z_2(\omega)=0$, Eq.~(\ref{eq:Final_Rel}) yields
\begin{align}
\label{eq:Env_Imps_case2A}
\Re\mbox{e}\left[Z_\textrm{eff}(\omega)\right]  \approx  \frac{R}{(C_\Sigma/C_g)^2 + (2 \omega R C)^2} \ ,
\end{align}
for a purely resistive environment $Z_g(\omega) = R$ and a symmetric turnstile $Z_{C_S}=Z_{C_D}=Z_{C}$.


In the particular case where $\Re\mbox{e}\left[Z_\textrm{eff}(\omega)\right]$ can be approximated with a frequency independent resistance $R$, Eq.~(\ref{eq:Rate_final}) becomes
\begin{eqnarray}
\label{eq:Rate_highT}
\Gamma_\textrm{AR}^\textrm{ hT} & \approx & \gamma_\textrm{env}^\textrm{D} \ \frac{\Delta^2}{(2 \pi)^4} \  \frac{\Delta}{\hbar \mathcal{N}}\left( \frac{R_K}{R_T} \right)^{\!\!\! 2} \int_{-\infty}^{0} d\xi_{\mathbf{k}_1} \ d\xi_{\mathbf{k}_2} \int_\Delta^{+\infty} d\varepsilon_{\mathbf{p}} \ d\varepsilon_{\mathbf{p'}} \nonumber \\
& \times & \left( \sqrt{\varepsilon_{\mathbf{p}}^2-\Delta^2} \ \sqrt{\varepsilon_{\mathbf{p'}}^2-\Delta^2}\right)^{-1} \ \left( S_{\mathbf{p}} \ S_{\mathbf{p'}}^* \right) /  \left( \xi_{\mathbf{k}_1,\mathbf{k}_2}^\textrm{c} \right)^{2} \ ,
\end{eqnarray}
in the high-temperature limit, $k_B T_\textrm{env} \gg  E_D^\textrm{out}(1,2) = 4E_C n_g - \Delta $ with $(1/4) \lesssim n_g \lesssim (3/4)$, assuming that the system is working at the optimal point $eV = \Delta$, and for large charging energy $E_C > \Delta$. In Eq.~(\ref{eq:Rate_highT}), we introduced the high-temperature Dynes parameter $\gamma_\textrm{env}^\textrm{D} = 2\pi (R/R_K)(k_BT_\textrm{env}/\Delta)$,~\cite{Pekola:2010,DiMarco} which is the only term of $\Gamma_\textrm{AR}^\textrm{ hT}$ which depends on $R$ and $T_\textrm{env}$.

\subsection{Numerical Results}

Using Eq.~(\ref{eq:Env_Imps_case2A}), the numerical integration of Eq.~(\ref{eq:Rate_final}) is relatively straightforward. Figure~\ref{fig:Rates_diff_T} shows the photon-assisted Andreev rate, Eq.~(\ref{eq:Rate_final}), as a function of the gate-induced charge $n_g$, for a single-electron transistor biased at the optimal voltage, $eV=\Delta$, and with charging energy $E_C>\Delta$. Each curve is obtained for different values of the temperature of the environment $T_\textrm{env}$. The other parameters are fixed to the values of sample S3 of Ref.~\onlinecite{Aref:2011}, as indicated in the figure. We see that the probability to have the tunneling of a Cooper-pair can be different from zero also away from the two-particle tunneling threshold, unlike the case without environment. In particular, the exchange of energy with the thermal bath in which the SET is embedded can make the Andreev reflection relevant even around the single-particle threshold. As a result, although the boundary of the Coulomb diamond corresponding to the transition $1 \rightarrow -1$ is avoided by means of the loop of Fig.~\ref{fig:diamonds_andreev2}, a Cooper pair can tunnel through the barrier of the drain, while $n_g$ goes back to $n_{g,1}$, before crossing the $1 \rightarrow 0 $ line. The decrease of $T_\textrm{env}$ leads to smaller values of $\Gamma_\textrm{AR}^\textrm{env}$ [see Fig.~\ref{fig:Rates_diff_T}], as well as the use of an electromagnetic environment with a smaller resistance $R$ [see Fig.~\ref{fig:Rates_diff_R}]. Whereas, in the latter case, the whole Andreev rate curve is shifted down proportionally to the ratio between initial and final resistances, the modulus of the first derivative of Eq.~(\ref{eq:Rate_final}) for $n_g > \Delta / 4 E_C$ increases proportionally to $T_\textrm{env}$ [see Fig.~\ref{fig:Rates_diff_T}].

The dependence of the photon-assisted Andreev rate, Eq.~(\ref{eq:Rate_final}), on the charging energy $E_C$ is shown in Fig.~\ref{fig:Rates_diff_Ec}. The increase of the ratio $E_C/\Delta >1$ allows to reduce the effect of the two-particle tunneling on the total electric current sustained by the SINIS turnstile. In particular, the lower is the effective temperature of the environment with respect to the critical temperature of the superconductor, the larger is the reduction of $\Gamma_\textrm{AR}^\textrm{env}$ upon increasing $E_C/\Delta$. Notice that the main effect of the change of the charging energy $E_C$ is the shifting of the environment-assisted Andreev rate along the induced-gate charge axis by the difference between the initial and final inverse ratios $\Delta / 4 E_C$.

Assuming that the number of electrons of the metallic island of the circuit of Fig.~\ref{fig:circuitBIS} decreases because of the tunneling of quasi-particles and Cooper pairs only, the total rate can be written as
$$
\Gamma_\textrm{tot} \simeq  \Gamma_{1 \rightarrow 0}^\textrm{\tiny{Dynes}} + 2 \Gamma_\textrm{AR}^\textrm{env}  \  .
$$
As a result, the error $\varepsilon_\textrm{acc} \equiv 2 \Gamma_\textrm{AR}^\textrm{env} / \Gamma_{1 \rightarrow 0}^\textrm{\tiny{Dynes}}$ determines how much the environment-assisted Andreev reflection affects the charge transport in the SINIS transistor. In particular, the condition $\varepsilon_\textrm{acc} < 10^{-8}$ is required for the metrological applications.~\cite{Flowers:2004} Figure~(\ref{fig:Ratios}) shows the ratio $\varepsilon_\textrm{acc}$ obtained from a numerical evaluation of Eqs.~(\ref{eq:Rate_final}) and (\ref{eq:single_rate}), as a function of $n_g$, when Eq.~(\ref{eq:Env_Imps_case2A}) holds. We see that $\varepsilon_\textrm{acc}$ is a non-monotonic function of $n_g$. Starting from the two-particle threshold occurring for $n_g=\Delta/4E_C$, this error first decreases exponentially as $n_g$ is increased. Then, close to the single-particle threshold, it rises up again reaching a local maximum value around $n_g=1/2-\Delta/4E_C$. For larger $n_g$ it tends exponentially to zero. Because of this kind of behavior, $\varepsilon_\textrm{acc}$ can be smaller or of the order of $10^{-8}$ when $\Delta/4E_C < n_g < 1/2-\Delta/4E_C $, and, at the same time, much larger than the value required by metrology around the single-particle threshold. Consequently, the time spent by the signal used to drive $n_g$ around $n_g=1/2-\Delta/4E_C$ has to be as small as possible in order to minimize the environment-assisted Andreev reflection.

From the experimental point of view, the determination, with a relatively high accuracy, of the values of the effective parameters of the environment, $R$ and $T_\textrm{env}$, is a tough challenge. The use of the Dynes parameter $\gamma_\textrm{\tiny{Dynes}}$, which in general depends also on $R$ and $T_\textrm{env}$, is preferred because it can be determined from the measured current-voltage characteristic of the SINIS turnstile. In this regard, the high-temperature two-particle tunneling rate Eq.~(\ref{eq:Rate_highT}) allows to study the photon-assisted Andreev reflection in terms of $\gamma_\textrm{\tiny{Dynes}}$ only. In Fig.~\ref{fig:RatiosHighT}, we plot the error $\varepsilon_\textrm{acc}$ obtained using Eq.~(\ref{eq:Rate_highT}) as a function of $n_g$. We see that the Dynes parameter, which typically ranges from $10^{-4}$ to $10^{-7}$, strongly affects $\Gamma_\textrm{AR}^\textrm{ hT}$ in the range $\Delta/4E_C < n_g < 1/2-\Delta/4E_C $. On the contrary, $\gamma_\textrm{\tiny{Dynes}}$ plays a minor role in the reduction of $\varepsilon_\textrm{acc}$ when $n_g$ is close to the single-particle threshold.

\begin{figure}[ht!]
\includegraphics[height=5.6cm]{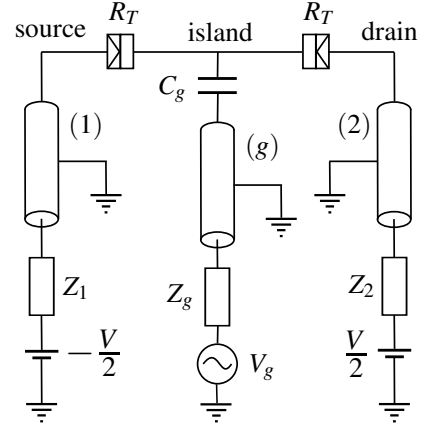}
\caption{Circuit representation of the hybrid S-I-N-I-S single-electron transistor (SET) connected to the impedances of the electromagnetic environment $Z_1(\omega)$, $Z_2(\omega)$, and $Z_g(\omega)$ by means of three transmission lines, (1), (2), and (g), respectively.}
\label{fig:circuit_TLs}
\end{figure}

\begin{figure*}[ht!]
\centering  \subfigure[]
{\includegraphics[height=5.4cm]{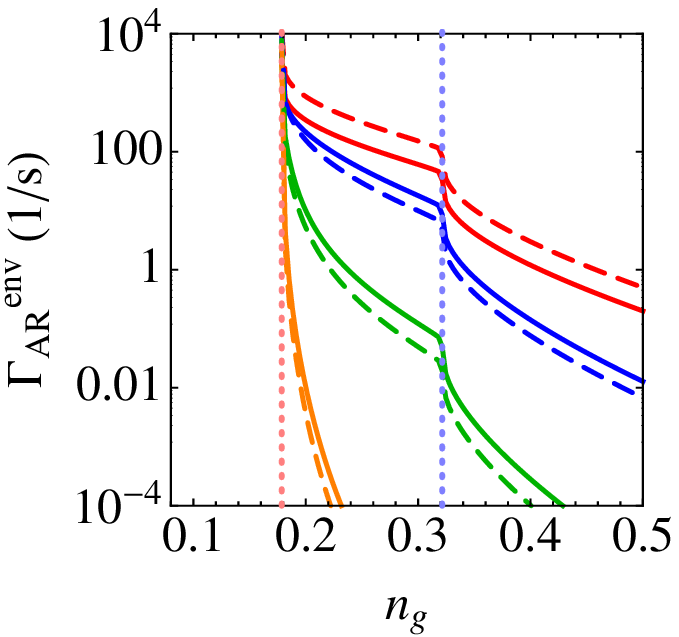}\label{fig:Rates_diff_R0_2B}}
\hspace{8mm} \subfigure[]
{\includegraphics[height=5.2cm]{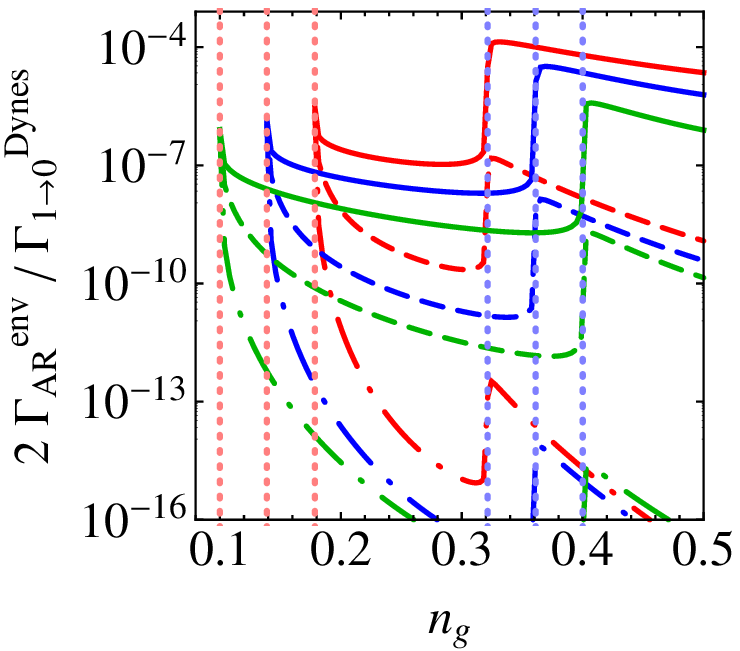}\label{fig:Ratios_diff_EcD_2B}}
\caption{(Color online) (a) Plot of the photon-assisted Andreev rate Eq.~(\ref{eq:Rate_final}) and (b) of the corresponding ratio $\varepsilon_\textrm{acc}$ as a function of the gate-induced charge $n_g$.  In panel (a), the solid and the dashed lines are obtained using Eqs.~(\ref{eq:Reff_TL_2B}) and (\ref{eq:Reff_TL_2B_approx}), respectively, setting $z_0=c_0=r_0=0$ (red lines) and using $z_0=0.7$ and $c_0=1$ with $r_0=5 \cdot 10^3$ (blue lines), $r_0=5 \cdot 10^4$ (green lines), and $r_0=5 \cdot 10^5$ (orange lines). In panel (b), the parameters for the plotted ratios are $z_0=c_0=r_0=0$ (solid lines) and $z_0=0.7$ and $c_0=1$ with $r_0=5 \cdot 10^4$ (dashed lines) and $r_0=5 \cdot 10^5$ (dotted-dashed lines). The curves with the same color are obtained using the same charging energy, $E_C/\Delta$: 1.4 with  $C=0.86 \cdot 10^{-16}$ (red lines), 1.8 with  $C=0.558 \cdot 10^{-16}$ (blue lines), 2.5 with  $C=0.262 \cdot 10^{-16}$ (green lines). In both panels, we set $\Delta=210\unit{\mu eV}$ (Aluminum), $R_T=430\unit{k\Omega}$, $C_g=10^{-16}$, $\mathcal{N}=100$, $\gamma_\textrm{\tiny{Dynes}}=10^{-5}$, $T_\textrm{env}=1.5\unit{K}$, and $R=10\unit{\Omega}$. The single- and two-particle thresholds, $1/2-\Delta/4E_C$ (light-blue vertical dotted lines), and $\Delta/4E_C$ (light-red vertical dotted lines), respectively, are also shown.}\label{fig:transLine}
\end{figure*}


\section{Effect of a resistive transmission line on the photon-assisted Andreev rate}

The results presented in the previous section have been obtained considering a SINIS turnstile directly connected to the external electromagnetic environment as illustrated in Fig.~\ref{fig:circuitBIS}. Using Eq.~(\ref{eq:Env_Imps_case2A}), we have shown that the smaller are the temperature $T_\textrm{env}$ and the resistance $R$ of the external circuit, as well as the total capacitance $C_\Sigma$, the lower is the Andreev tunneling rate Eq.~(\ref{eq:Rate_final}) with respect to the single-particle one. In particular, we have seen that the metrological accuracy may be reached for certain values of $T_\textrm{env}$, $R$, and $C_\Sigma$. However, in real experiments, the control of these parameters is typically limited. In general, their tuning to the desired values can be a difficult task. As discussed in Ref.~\onlinecite{DiMarco}, the insertion of cold and lossy transmission lines between the turnstile and the environment can help in overcoming this problem. One expects that such an indirect coupling allows a further reduction of the environment-assisted two-particle tunneling.

We therefore consider the circuit of Fig.~\ref{fig:circuit_TLs} where the three impedances of the environment are connected to the SINIS turnstile by means of three transmission lines. We assume that the latter are noiseless, i.e., at zero temperature. The noise across the drain of a SINIS device in such an indirect configuration is derived in Appendix \ref{app:Noise_Drain_TL}. In this case, the effective impedance $\Re\mbox{e}\left[Z_\textrm{eff}(\omega)\right]$ appearing in Eq.~(\ref{eq:Rate_final}) is given by Eq.~(\ref{eq:Final_Rel_TL_Fluct}). Let us assume again that $Z_1(\omega) \sim  Z_2(\omega) \sim  Z_g(\omega)$ and $C \sim C_g$ as well as that the three transmission lines have length $\ell$ and are all described by the same parameters $R_0$, $C_0$, and $L_0$, the resistance, the capacitance, and the inductance per unit length, respectively. Then, the three terms in the right-hand side of Eq.~(\ref{eq:Final_Rel_TL_Fluct}) contribute in a similar way to $\Re\mbox{e}\left[Z_\textrm{eff}(\omega)\right]$. In particular, they have the same order of magnitude for large $R_0$ and $\ell$. On the basis of these considerations and given the symmetry of the circuit of Fig.~\ref{fig:circuit_TLs}, we assume that the detrimental noise comes only from $Z_g(\omega)$, i.e., the voltage noises are $\delta V_g \neq 0$ and $\delta V_1 = \delta V_2 =0$, neglecting the effect of $Z_1(\omega)$ and $Z_2(\omega)$ as well as of the transmission lines (1) and (2). As a result, the effective impedance ``seen'' by the drain reduces to
\begin{equation}
\label{eq:Reff_TL_2B}
\Re\mbox{e}\left[Z_\textrm{eff}(\omega)\right]  \approx  R \ \left| \alpha_g(\omega) \right|^2 \ \left| T_g(\omega) \right|^2 \ ,
\end{equation}
with
\begin{align}
\label{eq:Alpha_g}
\alpha_g(\omega) = - \left[  2\frac{Z_{C_g}}{Z_C} \right. & + \frac{1}{2}\frac{Z_{C_g}}{Z_\infty^{(g)}}(\lambda_{C_g} + 1)  \nonumber \\
& + \left.  \lambda_g \left(  \frac{\lambda_{C_g}+1}{\lambda_{g}+1} \right)  e^{i K_g(\omega) \ell} \ T_g(\omega) \right]^{-1} \ ,
\end{align}
for $Z_g(\omega)=R$ and under the condition $Z_{1,2}[E_D^\textrm{out}(1,2)/\hbar] \ll \hbar / E_D^\textrm{out}(1,2) C_{S,D}$. The effective impedance Eq.~(\ref{eq:Reff_TL_2B}) tends to the asymptotic expression
\begin{equation}
\label{eq:Reff_TL_2B_approx}
\Re\mbox{e}\left[Z_\textrm{eff}(\omega)\right]  \approx  R \ \left(  \frac{C_g}{2C} \right)^2 \ \frac{e^{-\ell \sqrt{2\omega R_0 C_0}}}{1+\omega R_0 C_g^2 / C_0} \ ,
\end{equation}
if the transmission line is highly resistive, $R_0 \gg L_0 E_D^\textrm{out}(1,2) / \hbar$, and long enough, $\ell \sqrt{2 E_D^\textrm{out}(1,2) R_0 C_0 / \hbar} \gg 1$, and when the resistance of the environment is small, $R \ll R_0 C_g / 2 C_0$.~\cite{DiMarco} We see that Eq.~(\ref{eq:Reff_TL_2B_approx}) and in turn the environment-assisted Andreev rate Eq.~(\ref{eq:Rate_final}) decay exponentially in terms of $\ell$ and $R_0$.

In Fig.~\ref{fig:transLine}, we show the plots of the Andreev rates and the ratios $\varepsilon_\textrm{acc}$ resulting from the numerical integration of Eq.~(\ref{eq:Rate_final}) with $\Re\mbox{e}\left[Z_\textrm{eff}(\omega)\right]$ given by Eqs.~(\ref{eq:Reff_TL_2B}) and (\ref{eq:Reff_TL_2B_approx}). In both panels, we used the dimensionless parameters $z_0 = \sqrt{L_0/C_0}/R$, $c_0 = \ell C_0 / C$, and $r_0 = \ell R_0 / R$ whose values are chosen according to the analysis about the transmission function given in Ref.~\onlinecite{DiMarco} and in agreement with the currently achievable experimental values of $\ell$, $L_0$, $C_0$, and $R_0$. Similar results hold also in the configuration obtained restoring the fluctuations due to $Z_1(\omega)$ and $Z_2(\omega)$ and treating $Z_g(\omega)$ as a noiseless impedance.

From Fig.~\ref{fig:Rates_diff_R0_2B} we see that the bigger is $r_0$ the smaller is $\Gamma_\textrm{AR}^\textrm{env}$. In other words, a long and highly resistive transmission line allows a reduction of the environment-assisted Cooper-pair tunneling. By comparing the solid and the dashed lines obtained using Eqs.~(\ref{eq:Reff_TL_2B}) and (\ref{eq:Reff_TL_2B_approx}), respectively, we note that this decreasing is exponential-like. Eventually, the increase of $\ell$ and/or $R_0$ leads to a decrease of $\varepsilon_\textrm{acc}$ below $10^{-8}$, even close to the single-particle threshold [see Fig.~\ref{fig:Ratios_diff_EcD_2B}].
As a result, the use of a highly resistive and noiseless transmission line allows to filter out effectively the photon-assisted Andreev tunneling and, in particular, to reach the accuracy needed for metrological applications.

\section{Conclusions}
\label{sec:conclusions}

In this paper, we studied the environment-assisted Cooper-pair tunneling in a SINIS turnstile working in the Coulomb blockade regime. Specifically, we derived the Andreev reflection rate when only a single photon of the thermal bath is involved in the process. We found that the single-photon absorption enhances the two-electron tunneling from N to S. In particular, the probability per unit of time to have Andreev events is different from zero even for values of the induced gate charge $n_g$ close to the single-particle threshold $1/2 - \Delta / 4E_C$. As a result, the single-electron current, which is expected to be the dominant one in the device when $n_g$ follows the loop shown in Fig.~\ref{fig:diamonds_andreev2}, is also affected by the tunneling of Cooper pairs due to the environment. The influence of this source of error on the total current can be reduced by decreasing the effective resistance $R$ and temperature $T_\textrm{env}$ of the environment or, equivalently, the Dynes parameter $\gamma_\textrm{\tiny{Dynes}}$. The achievement of the metrological accuracy is also possible with the increasing of the charging energy $E_C$ with respect to the superconducting energy gap $\Delta$. We finally show that using a cold and lossy transmission line to couple indirectly the environment with the SINIS turnstile allows to reduce further the probability to have two-electron tunneling events, especially when $n_g$ crosses the single-particle threshold while covering the optimal loop.

\acknowledgments 

The authors thank D. V. Averin for useful discussions. Financial support from the Marie Curie Initial Training Network (ITN) Q-NET (project no. 264034), from Institut Universitaire de France and from Aalto University's ASCI visiting professor program is gratefully acknowledged. The work was partially supported by the Academy of Finland through its LTQ (project no. 250280) COE grant (VFM and JPP), and the National Doctoral Program in Nanoscience, NGS-NANO (VFM).



\appendix 

\section{Caldeira-Leggett model and Fluctuation-Dissipation Theorem}
\label{app:Caldeira_FluctDiss}

According to the Caldeira-Leggett model, the impedance $Z(\omega)$ of an electric circuit can be modeled as an ensemble of infinite quantum harmonic LC oscillators with Hamiltonian
$$
\hat{H}_\textrm{env} = \sum_\lambda \left[ \frac{\hat{Q}_\lambda^2}{2C_\lambda} + \frac{1}{2} C_\lambda \omega_\lambda^2 \left( \frac{\hbar}{e} \hat{\varphi}_\lambda \right)^2 \right] \ .
$$
The charge $\hat{Q}_\lambda$ and phase $(\hbar/e) \hat{\varphi}_\lambda$ operators play the role of the momentum and position respectively of the particle/oscillator $\lambda$ with mass $C_\lambda$ and characteristic frequency $\omega_\lambda^2=1/L_\lambda C_\lambda$. Each oscillator $\lambda$ of the ensemble/environment affects both the charge $\hat{Q}$ and phase $\hat{\varphi}$ of the circuit. In particular, the total phase fluctuation $\hat{\varphi}_\textrm{env}$ of $\hat{\varphi}$ due to $Z(\omega)$ is given by the superposition of all the phases of the oscillators of the environment, i.e., $\hat{\varphi}_\textrm{env}=\sum_\lambda \hat{\varphi}_\lambda$. Since $\hat{\varphi}_\lambda$ is the position operator of a harmonic oscillator, $\hat{\varphi}_\textrm{env}$ can be written as
\begin{equation}
\label{eq:Phi_env}
\hat{\varphi}_\textrm{env} = \sum_\lambda \rho_\lambda \left(  \hat{c}_\lambda^\dag +  \hat{c}_\lambda \right) \ ,
\end{equation}
in terms of the creation $\hat{c}_\lambda^\dag$ and annihilation $\hat{c}_\lambda$ operators of one photon. In Eq.~(\ref{eq:Phi_env}), we introduced the coupling term $\rho_\lambda=(e/\hbar)\sqrt{\hbar / 2 C_\lambda \omega_\lambda}$. In the Heisenberg picture, $\hat{\varphi}_\textrm{env}$ depends explicitly on time, with $\hat{c}^\dag_\lambda(t)= e^{+i\omega_\lambda t} \ \hat{c}^\dag_\lambda$ and $\hat{c}_\lambda(t)= e^{-i\omega_\lambda t} \ \hat{c}_\lambda$. The first time-derivative of Eq.~(\ref{eq:Phi_env}) gives the fluctuating voltage operator
\begin{equation}
\hat{V}_\textrm{env}(t)=\frac{\hbar}{e} \ \frac{d \hat{\varphi}_\textrm{env}(t)}{dt}=\frac{\hbar}{e} \ \sum_\lambda\rho_\lambda \ i \omega_\lambda \left[ \hat{c}^\dag_\lambda(t)- \hat{c}_\lambda(t) \right] \ ,
\end{equation}
whose mean value over the eigenstates of $\hat{H}_\textrm{env}$ is zero. On the other hand, the voltage-voltage correlation function $\delta\hat{V}_\textrm{env}(t,0) \equiv \left\langle \left\{ \hat{V}_\textrm{env}(t),\hat{V}_\textrm{env}(0) \right\} \right\rangle$ is
\begin{equation}
\label{eq:Volt_Volt}
\delta\hat{V}_\textrm{env}(t,0) =  \ \left( \frac{\hbar}{e} \right)^{\!\!\!\! 2} \sum_{\lambda\lambda'} \rho_{\lambda} \rho_{\lambda'} \ ( i\omega_\lambda) \ ( i\omega_{\lambda'}) \  C_\lambda (t,0) \ ,
\end{equation}
with
\begin{equation}
\label{eq:Corr}
C_\lambda (t,0) \equiv \left\langle \left\{ \left[ \hat{c}^\dag_\lambda(t)- \hat{c}_\lambda(t) \right] , \left[ \hat{c}^\dag_\lambda (0) - \hat{c}_\lambda (0) \right]   \right\} \right\rangle \ .
\end{equation}
The symbols $\{ , \}$ and $\langle \dots \rangle$ in Eq.~(\ref{eq:Corr}) indicate the anti-commutator and quantum mean value over the eigenstates of $\hat{H}_\textrm{env}$, respectively. Assuming that the number of photons of the environment is infinite, the terms in Eq.~(\ref{eq:Volt_Volt}) which create or destroy more than one photon can be neglected. Consequently, the correlation function $\delta\hat{V}_\textrm{env}(t,0)$ becomes
\begin{equation}
\label{eq:Corr2}
\delta\hat{V}_\textrm{env}(t,0) \ \simeq \ \left( \frac{\hbar}{e} \right)^{\!\!\!\! 2} \ \sum_{\lambda}\rho_\lambda^2 \ \omega_\lambda^2\left( e^{i\omega_\lambda t}+e^{-i\omega_\lambda t} \right) \left( 1+2n_\lambda \right) \ ,
\end{equation}
where $n_\lambda$ is the mean value of photons with frequency $\omega_\lambda$. The Fourier transform of Eq.~(\ref{eq:Corr2}) gives the spectral density function of the thermal bath,
\begin{eqnarray}
\label{eq:Corr_Fourier}
\left[ \delta\hat{V}_\textrm{env}(t,0) \right]_\omega \ & \simeq & \  \left( \frac{\hbar}{e} \right)^{\!\!\!\! 2} \ \sum_\lambda \rho_\lambda^2 \omega_\lambda^2 \ \coth\left(\frac{1}{2} \frac{\hbar\omega_\lambda}{k_B T_\textrm{env}} \right) \nonumber \\
& \times & 2 \pi \left[ \delta (\omega - \omega_\lambda )+\delta (\omega + \omega_\lambda ) \right]   \  .
\end{eqnarray}
To obtain Eq.~(\ref{eq:Corr_Fourier}) we assumed that $n_\lambda$ is given by the Bose-Einstein distribution function $n_{BE}(\omega_\lambda)=[\exp(\hbar\omega_\lambda/k_B T_\textrm{env})-1]^{-1}$ which satisfies the relation $1+2n_{BE}(x) = \coth(x/2)$. $T_\textrm{env}$ is the temperature of the environment.

On the other hand, assuming that the Fourier-transformed correlation function $\left[ \delta\hat{V}_\textrm{env}(t,0) \right]_\omega$ satisfies the quantum fluctuation-dissipation relation
\begin{equation}
\label{eq:fluct_diss}
\left[ \delta\hat{V}_\textrm{env}(t,0) \right]_\omega = 2 \hbar\omega \ \Re\mbox{e} \big[ Z(\omega) \big] \ \coth\left(\frac{1}{2} \frac{\hbar\omega}{k_B T_\textrm{env}} \right)  \  .
\end{equation} 
Comparing Eq.~(\ref{eq:Corr_Fourier}) with Eq.~(\ref{eq:fluct_diss}), we finally get the expression
\begin{equation}
\label{eq:fluctuation}
\Re\mbox{e} \left[ Z ( \omega ) \right] = \frac{R_K}{2}  \sum_\lambda \rho_\lambda^2 \omega_\lambda \left[ \delta (\omega - \omega_\lambda )+\delta (\omega + \omega_\lambda ) \right]  \  ,
\end{equation}
which allows to relate the macroscopic impedance $Z(\omega)$ with the microscopic quantities characterizing the environment.

\section{Voltage fluctuations across the drain}
\label{app:Noise_Drain}

\begin{figure}[hbtp]
\centering
\includegraphics[height=4.2cm]{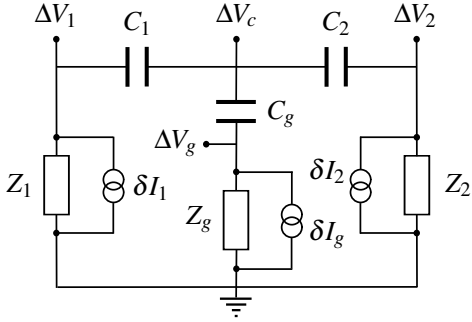}
\caption{Circuital scheme of a SINIS turnstile connected to an electromagnetic environment which produce current noise.}
\label{fig:circuit_noise}
\end{figure}

In this Appendix, we consider the circuit of Fig.~\ref{fig:circuit_noise}. We proceed in the evaluation of the voltage noise $\delta V = \Delta V_c - \Delta V_2$ across the capacitor $C_2$ and of its correlation function. The latter can refer, for instance, to the drain of the SINIS transistor.
%
%
%
%
%
%

We start by considering clockwise currents in the two meshes of the circuit of Fig.~\ref{fig:circuit_noise}. Then, assuming that the impedances $Z_1$, $Z_2$, and $Z_g$ produce the current noises $\delta I_1$, $\delta I_2$, and $\delta I_g$, respectively, the following equations
\begin{eqnarray}
\Delta I_1 & = & \delta I_1 - \frac{\Delta V_1}{Z_1} = - \frac{\Delta V_c - \Delta V_1}{Z_{C_1}} \ , \nonumber \\
\Delta I_2 & = & \delta I_2 + \frac{\Delta V_2}{Z_2} = \frac{\Delta V_c - \Delta V_2}{Z_{C_2}} \ , \nonumber \\
\Delta I_g & = & \delta I_g + \frac{\Delta V_g}{Z_g} = - \frac{\Delta V_c - \Delta V_g}{Z_{C_g}} \ , \nonumber \\
\Delta I_g & = & \Delta I_1 - \Delta I_2  \nonumber \ ,
\end{eqnarray}
hold. They can be rewritten as
\begin{eqnarray}
Z_{C_1} \delta V_1 & = & \left( Z_{C_1}  +  Z_1  \right) \Delta V_1 - Z_1 \Delta V_c \nonumber \\
Z_{C_2} \delta V_2 & = & - \left( Z_{C_2} + Z_2 \right) \Delta V_2 +  Z_2\Delta V_c \nonumber \\
Z_{C_g} \delta V_g & = &  - \left( Z_{C_g} +  Z_g \right) \Delta V_g + Z_g \Delta V_c \nonumber \\
0 & = & - Z_{C_g}Z_{c2}\Delta V_1 - Z_{C_g}Z_{c1}\Delta V_2 - Z_{C_1} Z_{C_2}\Delta V_g \nonumber \\
& + & \left( Z_{C_g}Z_{C_2} + Z_{C_g}Z_{C_1} + Z_{C_1} Z_{C_2} \right) \Delta V_c \nonumber \ ,
\end{eqnarray}
in terms of the voltage noises $\delta V_1=Z_1 \delta I_1$, $\delta V_2=Z_2 \delta I_2$, and $\delta V_g=Z_g \delta I_g$. By solving this system of equations one can get the unknown potentials $\Delta V_1$, $\Delta V_2$, $\Delta V_g$, and $\Delta V_c$. After some algebra, the voltage drop $\delta V$ reads as
\begin{equation}
\label{eq:voltage_noise}
\delta V = \frac{1}{\mathcal{Z}(\omega)}\left[ \mathcal{Z}_1(\omega) \delta V_1 + \mathcal{Z}_2(\omega) \delta V_2 - \mathcal{Z}_3(\omega) \delta V_g  \right]  \ ,
\end{equation}
where we introduced the impedances
\begin{align}
\label{eq:impedances_squared}
 \mathcal{Z}(\omega) & = \mathcal{Z}_3(\omega) [Z_2(\omega)+Z_{C_2}(\omega)] / Z_{C_2}(\omega) \nonumber \\
 & + \mathcal{Z}_1(\omega) [\mathcal{Z}_3(\omega) + Z_2(\omega) + Z_{C_2}(\omega)] / Z_{C_2}(\omega) \ , \nonumber \\
\mathcal{Z}_1(\omega) & = Z_g(\omega) + Z_{C_g}(\omega)  \ ,  \nonumber \\
\mathcal{Z}_2(\omega) & = Z_1(\omega) + Z_{C_1}(\omega) + Z_g(\omega) + Z_{C_g}(\omega) \ ,  \nonumber \\
\mathcal{Z}_3(\omega) & = Z_1(\omega) + Z_{C_1}(\omega) \ .
\end{align}
Here $Z_{C_j}(\omega) = i / ( \omega C_j )$ is the impedance of the $j$th capacitor with $j=1,2,g$. If $\delta V_1$, $\delta V_2$, and $\delta V_g$ satisfy the quantum fluctuation-dissipation theorem [see Eq.~(\ref{eq:fluct_diss})], then the voltage-voltage correlation function of $\delta V$ is
\begin{align}
\label{eq:fluctuation_correlation}
\left[ \delta\hat{V}(t,0) \right]_\omega & = 2\hbar\omega \ \left| \frac{\mathcal{Z}_1(\omega)}{\mathcal{Z}(\omega)} \right|^2 \ \Re\mbox{e}\left[ Z_1(\omega) \right] \coth\left(  \frac{1}{2} \frac{\hbar\omega}{k_B T_1} \right)  \nonumber  \\
& + 2\hbar\omega \ \left| \frac{\mathcal{Z}_2(\omega)}{\mathcal{Z}(\omega)} \right|^2 \ \Re\mbox{e}\left[ Z_2(\omega) \right] \coth\left(  \frac{1}{2} \frac{\hbar\omega}{k_B T_2} \right)  \nonumber  \\
& + 2\hbar\omega \ \left| \frac{\mathcal{Z}_3(\omega)}{\mathcal{Z}(\omega)} \right|^2 \ \Re\mbox{e}\left[ Z_g(\omega) \right] \coth\left(  \frac{1}{2} \frac{\hbar\omega}{k_B T_g} \right)  \  .
\end{align}
In Eq.~(\ref{eq:fluctuation_correlation}), we introduced the temperatures $T_1$, $T_2$, and $T_g$ of the impedances $Z_1$, $Z_2$, and $Z_g$, respectively. Similarly to Appendix \ref{app:Caldeira_FluctDiss}, the function $\left[ \delta\hat{V}(t,0) \right]_\omega$ can be related to the microscopic properties of the environment acting on the capacitance $C_2$ by comparing Eqs.~(\ref{eq:Corr_Fourier}) and (\ref{eq:fluctuation_correlation}):
\begin{align}
\label{eq:Final_Rel}
\frac{R_K}{2}\sum_\lambda \rho_\lambda^2 \omega_\lambda \delta ( |\omega| - \omega_\lambda ) & = \left| \frac{\mathcal{Z}_1(\omega)}{\mathcal{Z}(\omega)} \right|^2 \ \Re\mbox{e}\left[ Z_1(\omega) \right] \nonumber \\
& + \left| \frac{\mathcal{Z}_2(\omega)}{\mathcal{Z}(\omega)} \right|^2 \ \Re\mbox{e}\left[ Z_2(\omega) \right]  \nonumber \\
& + \left| \frac{\mathcal{Z}_3(\omega)}{\mathcal{Z}(\omega)} \right|^2 \ \Re\mbox{e}\left[ Z_g(\omega) \right] \ .
\end{align}
To obtain Eq.~(\ref{eq:Final_Rel}) we imposed that the temperature of the environment is uniform, $T_1=T_2=T_g=T_\textrm{env}$.

%
%
%
%
%
%

\section{Voltage fluctuations across the drain in the presence of transmission lines}
\label{app:Noise_Drain_TL}

\begin{figure}[ht!]
\centering
   \includegraphics[height=6cm]{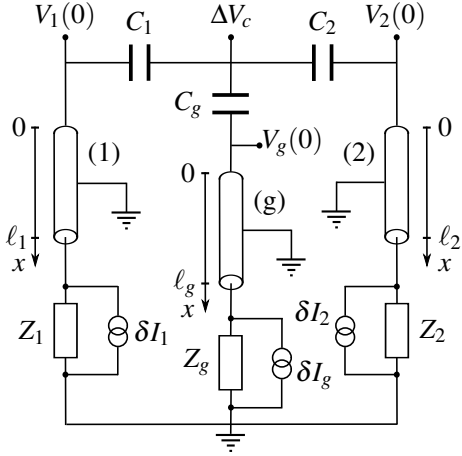}
\caption{Circuital scheme of a SINIS turnstile connected to a noisy electromagnetic environment by means of transmission lines.}
  \label{fig:circuit_noise_TL}
\end{figure}

In this Appendix, we analyze the circuit of Fig.~\ref{fig:circuit_noise_TL} where the transmission lines (1), (2), and (g) are inserted between the capacitances of the turnstile $C_1$, $C_2$, and $C_g$ and the effective impedances $Z_1$, $Z_2$, and $Z_g$ which give rise to the current noises $\delta I_1$, $\delta I_2$, and $\delta I_g$, respectively. The $j$th line, with $j=1,2,g$, has length $\ell_j$ and is characterized by the parameters $R_0^{(j)}$, $C_0^{(j)}$, and $L_0^{(j)}$, the resistance, the capacitance, and the inductance per unit length, respectively.
%
%
%
%
%
%
As in Appendix \ref{app:Noise_Drain}, we focus on the derivation of the voltage-voltage correlation function of the potential $\delta V = \Delta V_c - V_2(0)$ across $C_2$.

Imposing clockwise currents in the two main meshes of the circuit of Fig.~\ref{fig:circuit_noise_TL}, one can write the system of equations
\begin{eqnarray} 
\label{eq:system_Eqs_TL}
I_1(\ell_1) & = & \delta I_1 - \frac{V_1(\ell_1)}{Z_1} \ , \quad I_1(0)  =  - \frac{\Delta V_c -  V_1(0)}{Z_{C_1}} \ ,\nonumber \\
 \nonumber \\
I_g(\ell_g) & = & \delta I_g  + \frac{V_g(\ell_g)}{Z_g}  \ , \quad I_g(0)  =  \frac{\Delta V_c -  V_g(0)}{Z_{C_g}} \ , \nonumber \\
 \nonumber \\
I_2(\ell_2) & = & \delta I_2 + \frac{V_2(\ell_2)}{Z_2} \ , \quad I_2(0)  =  \frac{\Delta V_c -  V_2(0)}{Z_{C_2}} \ , \nonumber \\
I_g(0) & = & I_1(0) - I_2(0)  \ ,
\end{eqnarray}
where
\begin{align} 
\label{eq:V_and_I}
V_j(x) & = A_j e^{i K_j(\omega) x} + B_j e^{- i K_j(\omega) x} \ , \nonumber \\
I_j(x) & =  \left[ A_j e^{i K_j(\omega) x} - B_j e^{- i K_j(\omega) x} \right] / Z_\infty^{(j)} (\omega)
\end{align}
are the voltage and the current at a given point $x$ along the $j$th transmission line. In Eqs.~(\ref{eq:V_and_I}), we introduced the wave vector
$
K_j^2(\omega) = \omega^2 L_0^{(j)} C_0^{(j)} + i \omega R_0^{(j)} C_0^{(j)}
$
of the signal propagating along the $j$th line, and the impedance
$
Z_\infty^{(j)}(\omega) = i (R_0^{(j)} - i \omega L_0^{(j)}) / K_j(\omega)
$
[see also Refs.\onlinecite{DiMarco} and \onlinecite{Ingold:1992}]. The unknown coefficients $A_j$ and $B_j$ and the potential drop $\Delta V_c$ are the solutions of the system of equations presented in Eq.~(\ref{eq:system_Eqs_TL}). After some algebra, one can derive the voltage noise $\delta V$ across $C_2$:
\begin{equation}
\label{eq:Final_Rel_TL}
\delta V = \alpha_1(\omega) T_1(\omega) \delta V_1 + \alpha_g(\omega) T_g(\omega) \delta V_g + \alpha_2(\omega) T_2(\omega) \delta V_2 \ .
\end{equation}
Here $\delta V_1 = Z_1 \delta I_1$, $\delta V_2 = Z_2 \delta I_2$, and $\delta V_g = Z_g \delta I_g$. In Eq.~(\ref{eq:Final_Rel_TL}), we introduced the coefficients
\begin{equation}
\alpha_{1,g} (\omega) = - \frac{F(\omega)}{Z_{C_{1,g}}(\omega) Y(\omega)} \ , \quad \alpha_2 (\omega)  = 1 - \frac{F(\omega)}{Z_{C_2}(\omega) Y(\omega)} \ ,
\end{equation}
with
\begin{align}
F(\omega) & = \frac{1}{2}(1 - \lambda_{C_2}) + \lambda_{2} \left(  \frac{\lambda_{C_2}+1}{\lambda_{2}+1} \right)  e^{i K_2(\omega) \ell_2} T_2(\omega) \ , \nonumber \\
Y(\omega) & =  \frac{\lambda_{C_1}+1}{2 Z_\infty^{(1)}} - \lambda_{1} \left(  \frac{\lambda_{C_1}+1}{\lambda_{1}+1} \right)  e^{- i K_1(\omega) \ell_1} \ \frac{T_1(\omega)}{Z_{C_1}(\omega)} \nonumber \\
& + \frac{\lambda_{C_g}+1}{2 Z_\infty^{(g)}} + \lambda_{g} \left(  \frac{\lambda_{C_g}+1}{\lambda_{g}+1} \right)  e^{i K_g(\omega) \ell_g} \ \frac{T_g(\omega)}{Z_{C_g}(\omega)}  \nonumber \\
& + \frac{\lambda_{C_2}+1}{2 Z_\infty^{(2)}} +  \lambda_{2} \left(  \frac{\lambda_{C_2}+1}{\lambda_{2}+1} \right)  e^{i K_2(\omega) \ell_2} \ \frac{T_2(\omega)}{Z_{C_2}(\omega)} \ ,
\end{align}
where
\begin{align}
T_1(\omega) & = \frac{1}{2} \ \frac{(1-\lambda_{C_1})(\lambda_1 + 1)}{\lambda_1 \lambda_{C_1} e^{- i K_1(\omega) \ell_1} -  e^{i K_1(\omega) \ell_1}} \ , \nonumber \\
T_2(\omega) & = \frac{1}{2}\ \frac{(1-\lambda_{C_2}) (\lambda_2 + 1)}{ e^{- i K_2(\omega) \ell_2} -  \lambda_2 \lambda_{C_2} e^{i K_2(\omega) \ell_2}} \ , \nonumber \\
T_g(\omega) & = \frac{1}{2} \ \frac{(1-\lambda_{C_g}) (\lambda_g + 1)}{ e^{- i K_g(\omega) \ell_g} - \lambda_g \lambda_{C_g} e^{i K_g(\omega) \ell_g}} \nonumber
\end{align}
are the transmission functions, and
$$
\lambda_j(\omega) = \frac{Z_\infty^{(j)}(\omega) - Z_j(\omega)}{Z_\infty^{(j)}(\omega) + Z_j(\omega)} \ , \quad \lambda_{C_j}(\omega) = \frac{Z_\infty^{(j)}(\omega) - Z_{C_j}(\omega)}{Z_\infty^{(j)}(\omega) + Z_{C_j}(\omega)}
$$
the reflection coefficients.\cite{DiMarco}

Assuming that $\delta V_1$, $\delta V_2$, and $\delta V_g$ satisfy the fluctuation-dissipation theorem [see Eq.~(\ref{eq:fluct_diss})], and that $Z_1$, $Z_2$, and $Z_g$ are at the same temperature, we can finally get the relation
\begin{align}
\label{eq:Final_Rel_TL_Fluct}
\frac{R_K}{2}\sum_\lambda \rho_\lambda^2 \omega_\lambda \delta ( |\omega| - \omega_\lambda ) & = \left| \alpha_1(\omega) \right|^2 \ \left| T_1(\omega) \right|^2 \ \Re\mbox{e}\left[ Z_1(\omega) \right] \nonumber \\
& + \left| \alpha_2(\omega) \right|^2 \ \left| T_2(\omega) \right|^2 \ \Re\mbox{e}\left[ Z_2(\omega) \right]  \nonumber \\
& + \left| \alpha_g(\omega) \right|^2 \ \left| T_g(\omega) \right|^2 \ \Re\mbox{e}\left[ Z_g(\omega) \right] \ .
\end{align}
by means of Eq.~(\ref{eq:Corr_Fourier}).

%
%
%

\bibliography{references}

\end{document}